\documentclass[a4paper,11pt]{article}
\usepackage{graphicx,amssymb,amstext,amsmath,mathabx,mathtools,esvect,ulem}

\usepackage{floatflt}
\usepackage{subcaption}
\usepackage[dvipsnames]{xcolor}

\definecolor{cbl}{rgb}{0,0,1}                % bleu\

\newcommand{\bc}{\begin{center}}
\newcommand{\ec}{\end{center}}
\def\ba#1{\begin{array}{#1}\displaystyle}
\newcommand{\ea}{\end{array}}

\newcommand{\beq}{\begin{equation}}
\newcommand{\eeq}{\end{equation}}
\newcommand{\beqa}{\begin{eqnarray}}
\newcommand{\eeqa}{\end{eqnarray}}

\newcommand{\bi}{\begin{itemize}}
\newcommand{\ei}{\end{itemize}}
\newcommand{\p}{\partial}

\newcommand{\bra}{\langle}
\newcommand{\ket}{\rangle}

\newcommand{\Res}{{\rm Res}}

\newcommand{\TTb}{{T \bar{T}}}

\newcommand{\bal}{\boldsymbol{\alpha}}

\newcommand{\bel}{\boldsymbol{\beta}}

\newcommand{\OCA}{\color{ForestGreen}}

\newcommand{\FS}{\color{purple_nice}}

\usepackage{cite}

\definecolor{purple_nice}{rgb}{0.4,0.2,0.7}
\definecolor{fuel_blue}{RGB}{42,162,185}
\definecolor{YInMn_blue}{RGB}{46, 80, 144}
\definecolor{ultramarine}{RGB}{63, 0, 255}
\definecolor{KLEIN_blue}{rgb}{0, 0.18, 0.65}
\usepackage[linktocpage=true]{hyperref}
\hypersetup{
    colorlinks=true,
    linkcolor=YInMn_blue,
    citecolor=ultramarine,
    filecolor=fuel_blue,
    urlcolor=KLEIN_blue,
}
\begin{document}

\begin{titlepage}
\title{Complete Minimal Form Factors for Irrelevant Deformations\\ of Integrable Quantum Field Theory}
\author{Fabio Sailis{\color{blue} {$^\diamondsuit$}}, Olalla A. Castro-Alvaredo{\color{red} {$^\heartsuit$}} and Stefano Negro{\color{green} {$^\clubsuit$}}}
\date{\small {\color{blue} {$^\diamondsuit$}}  {\color{red} {$^\heartsuit$}}  Department of Mathematics, City St George's, University of London,\\ 10 Northampton Square EC1V 0HB, UK\\
\medskip
{\color{green} {$^\clubsuit$}}  Department of Mathematics, University of York,
Heslington, York YO10 5DD, UK\\
\medskip
}
\maketitle
\begin{abstract}
In this paper, we present a method to compute the minimal form factors (MFFs) of diagonal integrable field theories perturbed by generalized $\TTb$ perturbations. Building on existing results by the same authors, these MFFs are constructed in such a way as not to allow for any free parameters, an issue that plagued previous solutions. The MFFs are derived from a generalization of the standard integral representation which has been used  for UV-complete theories since the birth of the form factor bootstrap program. By UV-complete we mean theories whose short-distance/high-energy limit is a local conformal field theory. Their asymptotics is characterized by exponential decay at large rapidities. By computing higher particle form factors, we find that any natural higher-particle solutions involve the cancellation of parts of the newly found MFF. We conclude that the assumption that the form factor equations, particularly the kinematic residue equation, remain unchanged in the presence of $\TTb$ perturbations, is too strong. There is a trade-off between having MFFs satisfying desirable analyticity and asymptotic properties and finding analytic solutions to the form factor equations, which is likely solved by nontrivial changes to the form factor equations, especially those where locality or semilocality of fields are essential assumptions.
\end{abstract}

\bigskip
\bigskip
\noindent {\bfseries Keywords:} Integrable Quantum Field Theory, Irrelevant Perturbations, Form Factors

\vfill

\noindent 
{\Large {\color{blue} {$^\diamondsuit$}}}fabio.sailis@city.ac.uk\\
{\Large {\color{red} {$^\heartsuit$}}}o.castro-alvaredo@city.ac.uk\\
{\Large {\color{green} {$^\clubsuit$}}}stefano.negro@york.ac.uk\\

\hfill \today
\end{titlepage}
\section{Introduction}
Deformations of 2D quantum field theory (QFT) via irrelevant operators of the $\TTb$ family \cite{Smirnov:2016lqw, Cavaglia:2016oda, Dubovsky:2012wk, Caselle:2013dra} have been extensively studied in the literature. They are interesting because, on the one hand, they preserve integrability, should that be present in the original theory and, on the other hand, the resulting, still integrable model, has new very interesting properties. For this reason, $\TTb$ perturbations and their generalisations have been intensely studied in 2D (integrable) quantum field theory \cite{Smirnov:2016lqw, Cavaglia:2016oda, Conti:2018jho, Conti:2018tca, Conti:2019dxg, Dubovsky:2023lza}, in the context of the ODE/IM correspondence \cite{Aramini:2022wbn, Dorey:2019ngq}, via  thermodynamic Bethe ansatz (TBA) \cite{Hernandez-Chifflet:2019sua, Camilo:2021gro, Cordova:2021fnr, LeClair:2021opx, LeClair_2021, Ahn:2022pia, Dubovsky:2012wk, Caselle:2013dra}, perturbed CFT \cite{Guica:2017lia, Cardy:2018sdv, Cardy:2019qao, Aharony:2018vux, Aharony:2018bad, Guica:2020uhm, Guica:2021pzy, Guica:2022gts}, string theory \cite{Baggio:2018gct, Dei:2018jyj, Chakraborty:2019mdf, Callebaut:2019omt}, holography \cite{McGough:2016lol, Giveon:2017nie, Gorbenko:2018oov, Kraus:2018xrn, Hartman:2018tkw, Guica:2019nzm, Jiang:2019tcq,Jafari:2019qns}, quantum gravity \cite{Dubovsky:2017cnj, Dubovsky:2018bmo, Tolley:2019nmm, Iliesiu:2020zld, Okumura:2020dzb, Ebert:2022ehb}, out-of-equilibrium CFT \cite{Horvat2019,Medenjak:2020ppv,Medenjak:2020bpe,RMO}, in long-range spin chains \cite{Bargheer:2008jt,Bargheer:2009xy,PJG, Marchetto:2019yyt}, and employing the generalised hydrodynamics approach \cite{Doyon2022,Cardy:2020olv,Doyon2023}. A generalisation of this family of deformations has also been proposed for quantum-mechanical systems \cite{Gross:2019ach, Gross:2019uxi, Giordano_2023} and higher-dimensional field theories \cite{Bonelli:2018kik, Taylor:2018xcy, Conti:2022egv}.
The effect of the perturbation has been interpreted in multiple ways: as coupling the original QFT to two-dimensional topological gravity \cite{Dubovsky:2017cnj} or to random geometry 
\cite{Cardy:2018sdv} and as a state-dependent change of coordinates \cite{Conti:2018tca}.

Our own interest in this problem is motivated by integrable quantum field theory (IQFT) where many powerful techniques are at our disposal \cite{Zamolodchikov:1989hfa, Smirnov:2016lqw,Negro:2016yuu}. In particular, it is well known that in IQFT $\TTb$-like perturbations modify the exact two-body scattering matrix by a multiplicative (CDD) factor \cite{Castillejo:1955ed}. Throughout this paper, we will consider theories with a single particle spectrum. This means that the scattering matrix and scattering phase carry no particle indices. What follows can be easily generalised to theories with a richer spectrum. Then,  the deformed $S$-matrix is
\begin{equation}
    S_{\boldsymbol{\alpha}}(\theta) = \Phi_{\boldsymbol{\alpha}}(\theta) S(\theta)\;,
    \label{Smatrix}
\end{equation}
where $\boldsymbol{\alpha} = (\alpha_s)_{s\in\mathcal{S}\subset\mathbb{N}}$ is a, possibly infinite, vector, $S(\theta)$ is the two-particle scattering matrix for a theory with a single particle, and
\begin{equation}
    \Phi_{\boldsymbol{\alpha}}(\theta) = \exp\left[-i\sum_{s\in \mathcal{S}} \alpha_s m^{2s} \sinh(s\theta)\right]\;.
    \label{sum}
\end{equation}
Here $m$ is a fundamental mass scale such that the combination $\alpha_s m^{2s}$ is dimensionless. Hereafter, we will take $m=1$ for simplicity. $\mathcal{S}$ is a set of spin values, typically those of local conserved charges. Notice that $\mathcal{S}$ has to be a subset of the odd integers, otherwise the CDD factor does not satisfy the crossing equation $\Phi_{\boldsymbol{\alpha}}(i\pi-\theta) = \Phi_{\boldsymbol{\alpha}}(\theta)$. Since $\Phi_{\boldsymbol{\alpha}}(\theta)$ is a CDD factor, the theory described by this new $S$-matrix is still integrable and has the same particle spectrum as the original model.  
In \cite{GrosseGandalf_2007} certain theories of free fields $\phi(Q,x)$ on noncommutative Minkowski spaces with different noncommutativity parameters $Q$ were studied. These models were proven to satisfy a weaker form of localization called \textit{wedge localization}. Interestingly, these models can also be viewed as non-local (but wedge-local) field theories on a flat Minkowski space with a non-trivial S-matrix given precisely by the $s=1$ term in \eqref{sum}, i.e. the CDD factor coming from a ``pure" $\TTb$-deformation. The existence of non-trivial ``local" observables and their characterization is one of the main open questions in the study of these models. 

In our papers \cite{longpaper,ourTTb3,Castro-Alvaredo2024compl} we tackled the problem of computing matrix elements of ``local" operators via the form factor bootstrap program, the natural next step in the study of an IQFT once the $S$-matrix is known. Our study led to closed formulae for the form factors of a large class of fields and theories while also leaving several open questions. One of the most perplexing issues that we encountered is that the solutions of the form factor equations depend on a large number of free parameters (in \cite{longpaper,ourTTb3,Castro-Alvaredo2024compl} we called those $\beta_i$) and that there are no obvious physical requirements that allow us to fix these unambiguously. Another issue is that the factorized solutions that we found contain square roots, hence a degree of non-analyticity which stands at odds with the usual assumptions of the form factor program. Understanding the limit of validity of the standard techniques employed in the context of IQFT and their possible extension to these models is therefore one of the main objectives and challenges of our work.\\

To recap, let 
\beq 
F_n^{\mathcal{O}}(\theta_1\dots,\theta_n):=\bra 0| \mathcal{O}(0)|\theta_1, \dots \theta_n|0\ket=\bra 0| \mathcal{O}(0)|Z^{\dagger}(\theta_1), \dots , Z^{\dagger}(\theta_n)|0\ket\;,
\label{introFF}
\eeq 
be the $n$-particle form factor of an operator $\mathcal{O}$ between the ground state $|0\ket$ and a multiparticle in-state, consisting of $n$ particles of the same species and distinct rapidities $\theta_1 \ldots \theta_n$. The particle creation and annihilation operators, $Z^{\dagger}(\theta)$ and $Z(\theta)$, satisfy the Zamolodchikov-Faddeev algebra \cite{Zamolodchikov:1978xm,Faddeev:1980zy} 
\beqa 
Z(\theta_1)Z(\theta_2)&=&S(\theta_1-\theta_2)Z(\theta_2)Z(\theta_1)\;,\nonumber\\
Z(\theta_1)Z^\dagger(\theta_2)&=& S(\theta_2-\theta_1)Z^\dagger(\theta_2)Z(\theta_1) + \delta(\theta_1-\theta_2)\;,
\eeqa 
which generalizes to interacting theories the usual exchange relations of creation/annihilation operators in free models. As before, $S(\theta)$ is the two-body scattering matrix of a local IQFT. If the field is spinless the form factor depends only on rapidity differences. One of the more successful methods for computing these objects consists of solving a system of consistency equations \cite{weisz1977exact,KarowskiWeisz:1978vz,Karowski:1978eg,smirnov1992book} based on some natural physical assumptions such as unitarity, crossing symmetry, and the locality property of the field $\mathcal{O}$ being considered. These are:
\begin{itemize}
\item the \textit{braiding property}
    \beq 
    F_n^{\mathcal{O}}(\theta_1,\ldots,\theta_i,\theta_{i+1},\ldots,\theta_n)=S(\theta_{i}-\theta_{i+1}) F_n^{\mathcal{O}}(\theta_1,\ldots,\theta_{i+1},\theta_{i}, \ldots, \theta_n)\;,
    \label{W1}
    \eeq 
that descend directly from the exchange property of the Zamolodchikov-Faddeev algebra;
\item the \textit{monodromy property}
    \beq 
        F_n^{\mathcal{O}}(\theta_1+2\pi i , \theta_2 \ldots,\theta_n)= \gamma^\mathcal{O} F_n^{\mathcal{O}}(\theta_2, \ldots, \theta_n,\theta_1)\;,
        \label{W2}
    \eeq 
    
\item  the \textit{kinematical residue equation}
    \beq
        \lim_{\bar{\theta}\rightarrow \theta} (\bar{\theta}-\theta) F^{\mathcal{O}}_{n+2}(\bar{\theta}+i\pi, \theta, \theta_1,\ldots, \theta_n) = i \left(1-\gamma^{\mathcal{O}}\prod_{j=1}^n S(\theta-\theta_j)\right) F^{\mathcal{O}}_n(\theta_1,\ldots, \theta_n)\;.
        \label{KRE}
     \eeq
\end{itemize}
The factor of local commutativity $\gamma^\mathcal{O} $ first appeared in \cite{Yurov:1990} and encodes the (semi-)locality property of the fields. It is defined through the equal-time exchange relation of the operator $\mathcal{O}({\bf x})$ and the field $\phi({\bf y})$ associated with the particle creation operators $Z^{\dagger}(\theta)$. That is

\beq
\phi({\bf x}) \mathcal{O}({\bf y})= \gamma^\mathcal{O}\mathcal{O}({\bf y})\phi({\bf x}) \quad \textrm{for} \quad x^1>y^1\quad \mathrm{and} \quad x^0=y^0\;,
\label{eq:exch_rel}
\eeq
where formally
\beq
\phi({\bf x}) \sim \int d \,\theta \left[ Z(\theta) e^{-i {\bf p \cdot x}} + Z^{\dagger}(\theta)e^{i {\bf p \cdot x}} \right] \;.
\eeq
As an example, for the order/disorder operators in the Ising field theory, $\sigma$ and $\mu$, we have $\gamma^\mathcal{\sigma/\mu}= \pm 1 $. However, $\gamma$ can be a more general phase as it happens, for instance, in the Federbush model \cite{Feder1,Feder2,FRING_2002,Castro_Alvaredo_2001}. One can also have exchange relations more complicated than \eqref{eq:exch_rel}, where fields do not simply pick up a phase but become different fields, such as what happens with the branch point twist field in \cite{entropy}. A direct consequence is that equations (\ref{W2}) and (\ref{KRE}) cannot be satisfied if this factor is not included. In other words, the locality properties of the observables enter explicitly into the form factor equations and their understanding is pivotal for the computation of closed exact expressions  for $n$-particle form factors. If no bound states are present, (\ref{W1}\,--\,\ref{KRE}) are the only equations involved, and the solution procedure can be summarized in two steps: first one determines the simplest non-trivial form factor, the two-particle one; subsequently, the higher particle form factors are obtained by solving the recursive equation \eqref{KRE}.\\

In the case of a deformed model with $S$-matrix \eqref{Smatrix}, a two-particle form factor should satisfy
\beq 
F^{\mathcal{O}}_2(\theta;\bal)=S_{\bal}(\theta)F^{\mathcal{O}}_2(-\theta;\bal)=F^{\mathcal{O}}_2(2\pi i-\theta;\bal)\;.
\label{RHproblem}
\eeq 
Typically there is at least one other requirement, which specifies the residue of the form factor at the kinematic pole $\theta=i\pi$. However, if we limit our attention to solutions without poles, then we need to consider only the two equations \eqref{RHproblem}. Entire solutions to these equations are known as \emph{minimal form factors} (MFFs) and in the deformed theory we will denote these as $F_{\rm min}(\theta;\bal)$.
The same equations with $S_{\bal}(\theta)$ replaced by $S(\theta)$ are satisfied by the MFF of the undeformed theory, which we will denote as $F_{\rm min}(\theta)$:
\beq 
F_{\rm min}(\theta)=S(\theta)F_{\rm min}(-\theta)=F_{\rm min}(2\pi i-\theta)\;.
\label{FFE}
\eeq
This means that the MFFs $F_{\rm min}(\theta;\bal)$ and $F_{\rm min}(\theta)$ are proportional to each other through a function $\mathcal{D}_{\bal}(\theta)$ which satisfies
\beq
F_{\rm min}(\theta;\bal) =  \mathcal{D}_{\bal}(\theta) F_{\rm min}(\theta)\quad \Longrightarrow \quad \mathcal{D}_{\bal}(\theta)=\Phi_{\bal}(\theta) \mathcal{D}_{\bal}(-\theta)=\mathcal{D}_{\bal}(2\pi i-\theta)\;.
\label{DMM}
\eeq 
Note that each of the first equalities in (\ref{RHproblem}\,--\,\ref{DMM}) define a Riemann-Hilbert problem for the corresponding functions, with $S_{\bal}(\theta)$, $S(\theta)$ and $\Phi_{\bal}(\theta)$ playing the role of ``jump functions" \cite{gakhov2014boundary}. This can be used to explicitly write the solution as an integral representation involving the logarithm of the jump function. In fact, given the $S$-matrix of a massive, UV complete IQFT, it was shown in \cite{KarowskiWeisz:1978vz} that the MFF is uniquely fixed by the combined requirements of analyticity in the physical strip $\rm{Im}(\theta) \in [0,\pi]$ and (at most) exponential growth for $\theta$ large.
%and given by an integral formula, whose only input is the said $S$-matrix.  Again, this is a typical feature of a Riemann-Hilbert problem: the solution can be expressed by an integral representation whose only input is the logarithm of the jump function.
The integral representation assumes a particularly simple form in terms of the Fourier transform of the $S$-matrix phase:
\beq
i\delta(\theta):=\log{S(\theta)}= \int_0^{\infty}\frac{d\,t}{t} g(t) \sinh{\frac{t \theta}{i \pi}}
\label{Smatrixfourier}
\eeq
then
\beq
\rho(\theta):=\log{F_{\min}(\theta)}= \int_0^{\infty}\frac{d\,t}{t} \frac{g(t) }{\sinh{t}} \sin^2\frac{t(i\pi-\theta)}{2\pi}\,.
\label{KWFourierMFF}
\eeq
As we shall see below, the challenge for $\TTb$-perturbed theories lies precisely on finding convergent representations of this type.

For theories perturbed by $\TTb$ and its generalizations, it was found that the function $\mathcal{D}_{\bal}(\theta)$,  has the factorised structure \cite{Castro-Alvaredo2024compl,longpaper}
\beq 
\mathcal{D}_{\bal}(\theta):= \varphi_{\bal}(\theta)C_{\bel}(\theta)\;,
\label{minimal}
\eeq
with
\beq 
\varphi_{\bal}(\theta)=\exp\left[{\frac{
\theta-i\pi}{2\pi} \sum_{s\in \mathcal{S}} \alpha_s m^{2s} \sinh(s\theta)}\right]\;, \qquad 
C_{\bel}(\theta):=\exp\left[\sum_{n\in \mathbb{Z}^+} \beta_n m^{2n} \cosh(n\theta) \right]\;.
\label{betas}
\eeq 
Here the parameters $\alpha_s$ in (\ref{betas}) are those defining the CDD factor $\Phi_{\bal}(\theta)$ \eqref{sum}, while the parameters $\beta_n$ are, a priori, free and independent of the scattering phase. The existence of this very large freedom in the choice of the MFF is an ambiguity that was already highlighted in the original works \cite{ourTTb3,longpaper,Castro-Alvaredo2024compl} where -- arbitrarily and for simplicity -- it was chosen to set $\beta_n=0$ for all $n$. In a subsequent paper \cite{MMF} involving two of the present authors, the significance of the parameters $\beta_n$ was clarified. There, well-known integrable models such as the sinh-Gordon theory were interpreted as irrelevant perturbations of the Ising field theory involving an infinite set of fine-tuned\footnote{By fine-tuning here we mean that the resulting $S$-matrix determines a UV-complete IQFT. } couplings $\alpha_s$. Then, their MFF obtained from \eqref{Smatrixfourier}, \eqref{KWFourierMFF} can be rewritten in the form \eqref{minimal} where the ``free" parameters $\beta_n$ are now fixed in terms of the $\alpha_s$. It was then revealed that the role of the function $C_{\bel}(\theta)$ is to quell the unphysical asymptotic properties of the function $\varphi_{\bal}(\theta)$, producing a well-defined MFF that satisfies all physical requirements imposed by the UV-completeness of the theory. In the present context, UV-completeness means that the theory has a well-defined UV limit, described by a local CFT. A picture of IQFT as a massive, relevant perturbation of a CFT was famously put forward in \cite{Zamolodchikov:1989zs}.

The main result of this paper is a step by step procedure that allows us to fix the parameters $\beta_s$ for $\TTb$-deformed IQFTs in a physically and mathematically justified manner.
%
%In this paper we address the issue of fixing completely the MFF of $\TTb$-perturbed theories, when only a finite number of such perturbations are present.
We propose a generalisation of the integral representation that gives closed solutions for IQFTs with arbitrary $S$-matrices of the form (\ref{Smatrix}), (\ref{sum}), which
%the models under study and
reduces to the representation found in \cite{MMF} when the sets of couplings $\alpha_s$ is infinite and fine-tuned to yield a UV-complete theory.
%number of deformations is considered.
The parameters $\beta_s$ are fixed by analyticity and asymptotic requirements, but also, crucially, by the requirement that the resulting MFF is a smooth function of the perturbation parameters $\alpha_s$. For any set of parameters $\bal$ we find that the function $\mathcal{D}_{\bal}(\theta)$ is given by
\beq 
\mathcal{D}_{\bal}(\theta)=\mathcal{D}_{\bal}(i\pi)\prod_{s \in 2\mathbb{Z}^{+}-1}\left[-\frac{1}{2}\,e^{\frac{\theta-i\pi}{2}\sinh(s\theta)-\sum\limits_{n=1}^{s-1}\frac{\cosh(n\theta)}{s-m}+ c_{s}} \left(2i \sinh{\frac{\theta}{2}} \right)^{-\cosh{(s\theta)}}\right]^{\frac{\alpha_s}{\pi}}\;,
\label{original}
\eeq 
where $c_s$ is the constant (\ref{cn}).
This formula can be easily generalised to other interesting cases, such as the branch point twist fields (see \eqref{soltwist} in Appendix \ref{BPTFs}) and to the boundary case recently discussed in \cite{ourboundary}, where the deformation of the one-particle form factor is given by $\sqrt{\mathcal{D}_{\bal}(2\theta)}$. Note that in \eqref{original} and hereafter, we set the mass scale to $m=1$ to lighten the formulas.

\medskip
This paper is organised as follows: in Section \ref{section2} we give a heuristic derivation of our result, showing how generalised pairs $\delta(\theta)$-$\rho(\theta)$ can be constructed without relying on the existence of a Fourier transform $g(t)$. In Section \ref{section3} we present an alternative but equivalent derivation of the same result, starting from the standard contour-integral representation of the MFF proposed in the foundational paper \cite{KarowskiWeisz:1978vz} by Karowski and Weisz. In Section \ref{correlation} we discuss some properties of the correlation functions that might be obtained building on this new minimal form factor. 
We conclude in Section \ref{conclu}.
%We present two Appendices with details of our calculations.
In Appendix \ref{AppendixB} we discuss the construction of higher particle form factors starting from a factorised ansatz. In Appendix \ref{BPTFs} we generalise our construction of the MFF to branch point twist fields.

\section{A Heuristic Derivation: MFFs and their Fourier Transforms}
\label{section2}
In this section, we are going to produce a generalization of the integral representation of MFFs, capable of accounting for the presence of exponential CDD factors in the $S$-matrix. In fact, the representation \eqref{KWFourierMFF} relies on the possibility of expressing the scattering phase in integral form \eqref{Smatrixfourier}. This is not possible if the theory we are considering is a $\TTb$-deformed IQFT, whose $S$-matrix is of the form (\ref{Smatrix}) with (\ref{sum}). Our goal is then to knead \eqref{Smatrixfourier} and \eqref{KWFourierMFF} into a new, more general form. Note that the manipulations we will be performing here are formal and that we will be quite cavalier about convergence issues. As such, the content of this section is to be taken as a heuristic derivation of our main results.

\subsection{Integral Representation of the Phase Shifts}

Let us consider an interacting factorized scattering theory with diagonal $S$-matrix,
\begin{equation}
    S(\theta) = \epsilon\, e^{i \delta(\theta)}\;.
    \label{SmatrixExp}
\end{equation}
The quantity $\delta(\theta)$ is the \emph{phase shift} and is normalized so that
\begin{equation}
    \delta(0) = 0\;.
\end{equation}
Consequently, the remaining constant $\epsilon$ encodes the value of the $S$-matrix at zero rapidity difference $S(0)=\epsilon$. For $\epsilon=\pm 1$ we have the usual bosonic/fermionic statistics. However, more general theories exist, such as the Federbush model \cite{Feder1,Feder2} where $S(0)$ is a generic phase (in this example, there are also two particle types). If our factorized scattering theory is a UV complete IQFT, then the phase shift can be Fourier transformed as in \eqref{Smatrixfourier}. The converse is however not true. A good counterexample is the ``bosonic" sinh-Gordon model studied in \cite{bosonic} whose $S$-matrix has a well defined Fourier transform despite the theory lacking UV completion. More general cases have been discussed in \cite{Camilo:2021gro}. The meaning of this statement is made clearer below. 

\subsection{Asymptotic Behavior of the $S$-Matrix and Temperedness}

We now discuss in more detail the conditions for existence of the Fourier transform of the phase shift, that is the notion of {\it tempered distribution} \cite{GelfanShilovdistr,tempered}. We show that the possibility of defining the phase shift as a Fourier transform is related to its nice decaying properties at large rapidities and that $\TTb$-perturbed $S$-matrices are precisely of the type where this condition is violated.

Let $f$ be a function on $\mathbb{R}$ and consider $\mathcal{D} (\mathbb{R})$, the class of test functions on $\mathbb{R}$ and $\mathcal{D}' (\mathbb{R})$, the class of distributions on $\mathbb{R}$, i.e. the set of linear functionals on $\mathcal{D} (\mathbb{R})$. Explicitly, given $f:\mathbb{R}\to \mathbb{R}$ and $\varphi\in \mathcal{D} (\mathbb{R})$ it is possible to define a distribution\footnote{Note that this symbols $\mathcal{D}$ and $\mathcal{D}'$ have nothing to do with the function $\mathcal{D}_{\bal}(\theta)$ defined earlier (\ref{minimal}). Similarly, the symbols $\mathcal{S}$ and $\mathcal{S}'$ around equation (\ref{23}), are distinct from the set of spins $\mathcal{S}$ in (\ref{sum}).} 

\beq
\bra f, \varphi \ket =\int_{-\infty}^{\infty} f(x) \varphi(x) dx\, \in \mathcal{D}' (\mathbb{R}).
\label{distributiondefinition}
\eeq

Suppose that $\delta(\theta)$ is a smooth function satisfying the asymptotic condition
\beq
\lim_{|\theta| \to \infty} \delta(\theta) = 0.
\eeq
This means that this function and all its derivatives decay faster than any inverse polynomial, i.e.,
\beq
\forall n \in \mathbb{N}, \quad \sup_{\theta \in \mathbb{R}} \left| \theta^k \delta^{(n)}(\theta) \right| < \infty \quad \text{for all } k \in \mathbb{N},
\label{polynomialbound}
\eeq
Therefore $\delta(\theta)$ belongs to the Schwartz class $\mathcal{S}(\mathbb{R})$. This implies that its Fourier transform
\beq
g(t) := \int_{-\infty}^{\infty} \frac{d\theta}{2\pi} \, e^{i t \theta} \delta(\theta)
\label{23}
\eeq
exists as an ordinary Lebesgue integral and defines a function $g(t) \in \mathcal{S}'(\mathbb{R})\subseteq \mathcal{D}'(\mathbb{R})$, the dual to the Schwartz class, i.e. the class of \textit{tempered distributions}. In particular, we also have the inversion formula
\beq
\delta(\theta) = \int_{-\infty}^{\infty} dt \, e^{-i t \theta} g(t),
\eeq
with absolute convergence of the integral due to rapid decay of $g(t)$. In summary, if $\delta(\theta)\in L^1(\mathbb{R})$, meaning that its absolute value is Lebesgue-integrable, we can define its Fourier transform, which will also have nice decaying properties by construction. This is what happens for standard phase shifts. Now, consider a phase shift coming from a $\TTb$-deformation: $\delta(\theta) \sim\sinh(s \theta)$. It still defines a distribution in the sense of $\mathcal{D}'(\mathbb{R})$ (class of distributions in $\mathbb{R}$),
i.e., for instance, it acts continuously on test functions with compact support:

\beq
\varphi(\theta) \in C_c^\infty(\mathbb{R}) \quad \implies \quad \langle \sinh, \varphi \rangle = \int_{-\infty}^\infty \sinh(s \theta) \, \varphi(\theta) \, d\theta \quad \text{is finite}.
\label{distribuzione sinhx}
\eeq
However if we consider the real line, $\sinh(s\theta)$ grows exponentially as $|\theta| \to \infty$:
\beq
|\sinh(s\theta)| \sim \frac{1}{2} e^{s|\theta|},
\eeq
and therefore it does not satisfy any polynomial bound of the form \eqref{polynomialbound} which is a necessary condition for tempered distributions. Consequently, $\delta(\theta)$ in this case is a distribution, but it is not tempered; thus, its Fourier transform is not well defined within the classical theory of tempered distributions. Indeed, the subclass of test functions $\mathcal{E}(\mathbb{R})\subset \mathcal{D}(\mathbb{R})$ necessary to make an integral like \eqref{distribuzione sinhx} convergent, 
can  only be of two kinds:  the class of test functions with compact support $\mathcal{E}(\mathbb{R})=C_c^\infty(\mathbb{R})$ as written above or a subclass with a quicker exponential decay than $\exp{s|\theta|}$:
\beq
\mathcal{E}_{s}(\mathbb{R}) := \left\{ \varphi \in C^\infty(\mathbb{R}) \;\middle|\; \exists s > 0 \;\text{ such that }\; \sup_{\theta \in \mathbb{R}} \left| \varphi(\theta) e^{s |\theta|} \right| < \infty \right\} \subset \mathcal{D}(\mathbb{R})\,.
\eeq

\subsection{The Scattering Phase and its Fourier Transform}
Let us now apply these ideas to the integral representation of a generic $S$-matrix. Consider then the scattering phase $\delta(\theta)$ and suppose it can be written in terms of its Fourier transform as
\begin{equation}
    \delta(\theta) = \frac{i}{2} \int_{-\infty}^{\infty} \frac{dt}{t} g(t) e^{\frac{it\theta}{\pi}}\;.
\label{eq:KW_rep_phase_shift}
\end{equation}
Since the phase shifts are taken to be real functions of the rapidity difference, the function $g(t)$ must be even: $g(-t) = g(t)$. We can then rewrite the Fourier transform in the form (\ref{Smatrixfourier})
\begin{equation}
    \delta(\theta) =  -\int_{0}^{\infty} \frac{dt}{t} g(t) \sin\frac{t\theta}{\pi}\;,
\label{eq:KW_rep_phase_shift_2}
\end{equation}
which is the representation given earlier (\ref{Smatrixfourier}).
In order for this Fourier transform to be a proper function -- rather than a distribution -- we must demand that $g(t)$ approaches constant values at the integral bounds
\begin{equation}
    g(t) = g^{(0)} + O(t^2)\;,\qquad \lim_{\vert t\vert\to\infty} g(t) = g^{(\infty)} < \infty\;.
\end{equation}
Note that we assumed $g(t)$ to be Taylor expandable around $t=0$ and to possess only isolated singularities. The constants $g^{(0)}$ and $g^{(\infty)}$ may vanish.
We can think of $g(t)$ as a function on the Riemann sphere $\mathbb{C}\cup\{\infty\}$ which means that it must be either  a constant or a meromorphic function. In both cases, we can use a Mittag-Leffler expansion \cite{MiLe} to write
\begin{equation}
    g(t) = g^{(0)} - \sum_{n=1}^{\infty} \frac{t^2}{\pi p(n)} \frac{g^{(n)}}{t^2 + p(n)^2}\;,\qquad g^{(n)} = \pm 2 \pi i \underset{t = \pm i p(n)}{\rm Res}[g(t)]\;,
\label{eq:ML_expansion_1}
\end{equation}
for some set of poles $\{\pm i p(n) \,\vert\, n = 1,\ldots, \infty \}$. We can use the physical properties of the phase shifts to specify the positions of these poles. In particular, we want $\delta(\theta)$ to be $2\pi i$-periodic. We see that \eqref{eq:ML_expansion_1} and the residue theorem -- which we can use since the integrand of \eqref{eq:KW_rep_phase_shift} decays as $t^{-1}$ as $\vert t\vert \to \infty$ -- imply
\begin{equation}
    \delta(\theta) = {\rm const.} +  \sum_{n=1}^\infty \frac{g^{(n)}}{2 p(n)} e^{-\frac{p(n)\theta}{\pi}}\;,\quad \forall \, \theta>0.
\end{equation}
If we want this expression to be $2\pi i$-periodic, then we need to fix
\begin{equation}
    p(n) = n \pi\,.
\end{equation}
In conclusion, the function $g(t)$ must take the following form
\begin{equation}
    g(t) = g^{(0)} - \sum_{n=1}^\infty \frac{t^2}{n \pi^2} \frac{g^{(n)}}{t^2 + n^2 \pi^2}\;.
\label{eq:ML_expansion_2}
\end{equation}

\subsection{The Logarithm of the MFF and its Fourier Transform}
Given the integral representation \eqref{eq:KW_rep_phase_shift_2}, we introduce the new function
\begin{equation}
    \rho(\theta) = -\frac{1}{4} \intop_{-\infty}^\infty \frac{dt}{t} \frac{g(t) - g^{(0)}}{\sinh t} e^{\frac{t(i\pi-\theta)}{i\pi}}  + \frac{g^{(0)} - 1 + \epsilon}{2} \log\left(-i\sinh\frac{\theta}{2}\right)\;.
\label{eq:KW_rep_min_ff_1}
\end{equation}
Note that the Fourier integral converges as long as $0 \leq {\rm Im}\,\theta \leq 2\pi$. We immediately verify the identities\footnote{For equation (\ref{eq:min_ff_eq_rho_2}), we can use the known integral $\int_{-\infty}^{\infty} \frac{dt}{t}\sin\frac{t \theta}{\pi} = \pi \,{\rm sign}(\theta)$ and 
\beq 
\log(-i\sinh\frac{\theta}{2}) - \log(i \sinh\frac{\theta}{2}) = -i\pi \, {\rm sign}(\theta)\,.
\label{25}
\eeq }
\begin{subequations}
    \begin{align}
        & \rho(i \pi - \theta) = \rho(i\pi + \theta)\;, \label{eq:min_ff_eq_rho_1} \\
        & \rho(\theta) - \rho(-\theta) = i \delta(\theta) + \frac{i\pi (1-\epsilon)}{2}\,{\rm sign}(\theta)\;. \label{eq:min_ff_eq_rho_2}
    \end{align}
\label{eq:min_ff_eq_rho}
\end{subequations}
Consequently, we can take this function to be the logarithm of the MFF
\begin{equation}
    \log F_{{\rm min}}(\theta) = \rho(\theta) \quad \Longrightarrow \quad  \begin{cases} F_{{\rm min}}(\theta) = F_{{\rm min}}(2\pi i - \theta)\;, \\
    F_{{\rm min}}(\theta) = S(\theta) F_{{\rm min}}(-\theta)\;.
    \end{cases}
\end{equation}
With some manipulations\footnote{We use another known integral: $\int_{-\infty}^{\infty} \frac{dt}{t} \frac{\sin^2(t x)}{\sinh t} = \frac{1}{2}\log\cosh(\pi x)$.} we can rewrite $\rho(\theta)$ as follows
\begin{equation}
    \rho(\theta) = \rho(i\pi) + \intop_0^\infty \frac{dt}{t} g(t) \frac{\sin^2\left(t \frac{i\pi-\theta}{2\pi}\right)}{\sinh t} - \frac{1-\epsilon}{2}\log\left( - i \sinh\frac{\theta}{2}\right) \;,
\end{equation}
where $\rho(i\pi)$ is a constant that can be absorbed into the normalisation of the MFF.

\subsection{From Integral to Series Representations}
\label{mittag}
 Using the Mittag-Leffler expansion \eqref{eq:ML_expansion_2} in the integral representation \eqref{eq:KW_rep_phase_shift}, we can write
\begin{equation}
    \begin{split}
        \delta(\theta) & = - g^{(0)} \intop_0^\infty \frac{dt}{t} \sin \frac{t\theta}{\pi} + \sum_{n=1}^\infty \frac{g^{(n)}}{n\pi} \frac{1}{2\pi i} \int_{-\infty}^\infty dt \frac{t\,  e^{ \frac{i t \theta}{\pi}}}{t^2 + n^2 \pi^2} \\
        & = \left[-\frac{\pi g^{(0)}}{2} + \frac{1}{2\pi} \sum_{n=1}^\infty \frac{g^{(n)}}{n} e^{-n \vert \theta \vert}\right] {\rm sign}(\theta)\;.
    \end{split}
\label{eq:delta_series}
\end{equation}
Similarly, putting together the integral representation (\ref{eq:KW_rep_min_ff_1}) and the expansion (\ref{eq:ML_expansion_2}), we get
\begin{equation}
        \rho(\theta) = \frac{i}{2\pi} \sum_{n=1}^\infty \frac{g^{(n)}}{n} \frac{1}{2\pi i}\intop_{-\infty}^\infty dt \frac{t}{t^2 + n^2 \pi^2} \frac{e^{\frac{t(i\pi-\theta)}{i\pi}} }{\sinh t} + \frac{g^{(0)} - 1 + \epsilon}{2}\log \left(-i\sinh\frac{\theta}{2}\right)\,.
\label{eq:rho_series1}
\end{equation}
The integral can be computed by residue theorem, with contributions from a double pole at $t=i\pi n$ and simple poles at $t=i\pi m$ with $m\neq n$. This gives\footnote{Note that this integral needs regularization: we suppose that $\theta$ has a vanishingly small imaginary part.}
\beqa 
\frac{1}{2\pi i} \intop_{-\infty}^\infty dt \, \frac{t}{t^2 + n^2 \pi^2} \frac{e^{\frac{t(i\pi-\theta)}{i\pi}} }{\sinh t} & \underset{\theta > 0}{=} & \sum_{m=1}^\infty \underset{t = i m \pi}{\rm Res}\left[\frac{t}{t^2 + n^2 \pi^2} \frac{e^{\frac{t(i\pi-\theta)}{i\pi}} }{\sinh t}\right]\nonumber\\
&=& i \frac{2n\pi\, 
        \left(\frac{\theta-i\pi}{\pi}\right) - 1}{4 \pi n} e^{-n \theta} + \frac{i}{\pi}\sum_{\genfrac{}{}{0pt}{}{m=1}{m\neq n}}^\infty \frac{m\,  e^{-m \theta}}{n^2 - m^2}\,.
\eeqa  
For generic $\theta$ the formula above generalised in a simple way, which can be captured by introducing the ${\rm sign}(\theta)$ function in the appropriate place, giving the final result:

\begin{equation}
    \begin{split}
        \rho(\theta) 
        & = -\frac{1}{2\pi^2}\sum_{n=1}^\infty \frac{g^{(n)}}{n}\left[\frac{2n\pi\, {\rm sign}(\theta)
        \left(\frac{\theta-i\pi}{\pi}\right) - 1}{4 n} e^{-n\vert \theta\vert} + \sum_{\genfrac{}{}{0pt}{}{m=1}{m\neq n}}^\infty \frac{m\, e^{-m\vert \theta\vert}}{n^2 - m^2}  \right] \\
        &\phantom{= \qquad} + \frac{g^{(0)} - 1 + \epsilon}{2}\log \left(-i\sinh\frac{\theta}{2}\right) \;.
    \end{split}
\label{eq:rho_series}
\end{equation}
Let us compute the difference $\rho(\theta) - \rho(-\theta)$ and check that (\ref{eq:min_ff_eq_rho_2}) is satisfied. Thanks to the relation (\ref{25}) we have that the last term in \eqref{eq:rho_series} produces the first term in \eqref{eq:delta_series}. 
Furthermore, the sum over $m$ in \eqref{eq:rho_series} is symmetric under $\theta\to-\theta$, so it does not contribute and, finally, the remaining term in \eqref{eq:rho_series} yields the sum over $n$ in \eqref{eq:delta_series}:
\begin{equation}
    \frac{2n\pi\, {\rm sign}(\theta)\left(\frac{\theta-i\pi}{\pi}\right) - 1}{4 n} e^{-n\vert \theta\vert} - \frac{2n\pi\, {\rm sign}(-\theta)\left(\frac{-\theta-i\pi}{\pi}\right) - 1}{4 n} e^{-n\vert -\theta\vert} = -i\pi\, {\rm sign}(\theta) e^{-n\vert \theta\vert}\;.
\end{equation}
\subsection{Recovering a $\TTb$-Like Representation}
So far, the contents of this section provide a rederivation of known results. We now come to the new conceptual step which will allow us to use these results to construct a pair $\delta(\theta)$-$\rho(\theta)$ which corresponds to IQFTs perturbed by $\TTb$. The key observation is that, for the purpose of the above identities, the only property that matters is the parity of the functions $e^{-n\vert\theta\vert}$ and ${\rm sign}(\theta) \theta e^{-n\vert\theta\vert}$, that is, that the former is even under $\theta\to -\theta$, while the latter is odd. This means that we could replace these functions with others having the same parity properties and the MFF equations \eqref{eq:min_ff_eq_rho} would still hold. We will now exploit this property by investigating what the most general choices for these functions can be. 
We perform the substitutions 
\begin{equation}
    e^{-n\vert\theta\vert} \,{\rm sign}(\theta) \;\longrightarrow \; f_n(\theta)\;,\qquad e^{-n\vert\theta\vert} \theta \,{\rm sign}(\theta) \;\longrightarrow \; g_n(\theta)\;,\qquad e^{-n\vert\theta\vert} \;\longrightarrow \; h_n(\theta)\;.
\end{equation}
in the series expansions (\ref{eq:delta_series}) and (\ref{eq:rho_series}). 
For simplicity, we will also take $g^{(0)} = 0$ and $\epsilon=1$ so that the term $\log(-i\sinh\frac{\theta}{2})$ in (\ref{eq:rho_series}) drops out and we have an $S$-matrix which is a ``pure" $\TTb$ deformation.  We then have
\begin{equation}
    \begin{split}
        \delta(\theta) & = \frac{1}{2\pi} \sum_{n=1}^\infty \frac{g^{(n)}}{n} f_n(\theta)\;, \\
        \rho(\theta) & = -\frac{1}{2\pi^2} \sum_{n=1}^\infty \frac{g^{(n)}}{n} \left[ - \frac{\pi i}{2} f_n(\theta) + \frac{1}{2} g_n(\theta) - \frac{1}{4n} h_n(\theta) + \sum_{\genfrac{}{}{0pt}{}{m=1}{m\neq n}}^\infty \frac{m}{n^2 - m^2} h_m(\theta)\right]\;.
    \end{split}
\label{eq:delta_rho_ansatz}
\end{equation}
We want these two quantities to be, respectively, a phase shift and the logarithm of the corresponding MFF. Clearly, only some functions $f_n(\theta)$, $g_n(\theta)$ and $h_n(\theta)$ will work. We must have
\begin{equation}
    \begin{split}
        \delta(\theta) + \delta(-\theta) = 0\;& ,\qquad \textrm{unitarity of the } S\textrm{-matrix}\;, \\
        \delta(\theta) - \delta(i\pi - \theta) = 0\;& ,\qquad \textrm{crossing symmetry of the }S\textrm{-matrix}\;, \\
        \rho(\theta) - \rho(-\theta) = i \delta(\theta)\;& ,\qquad \textrm{Watson's equation}\;, \\
        \rho(i\pi + \theta) - \rho(\pi i -\theta) = 0\;& ,\qquad \textrm{crossing symmetry of the form factor}\;.
    \end{split}
\label{eq:delta_rho_conditions}
\end{equation}
We then see immediately that the functions $f_n(\theta)$ have to satisfy the conditions
\begin{equation}
    f_n(\theta + i \pi) = f_n(-\theta) = - f_n(\theta) = -f_n(\theta - i\pi)\;.
\label{eq:eqf}
\end{equation}
Watson's equation tells us that $g_n(\theta)$ and $h_n(\theta)$ have to be even 
\begin{equation}
    g_n(-\theta) = g_n(\theta)\;,\qquad h_n(-\theta) = h_n(\theta)\;,
\label{eq:eqgh}
\end{equation}
and, finally, the crossing symmetry of the form factor yields the following identities\footnote{Here we are assuming that each term in the series -- also in the double series -- appearing in the equations \eqref{eq:delta_rho_conditions} has to vanish \emph{independently}.}
\begin{equation}
    g_n(\theta + i \pi) - g_n(\theta - i \pi) = 2 \pi i f_n(\theta + i \pi)\;,\qquad h_n(\theta + i \pi) = h_n(\theta - i \pi)\;.
\label{eq:eqgh2}
\end{equation}
If we require that $\{f_n(\theta)\}_{n\in\mathbb{Z}^+}$, $\{g_n(\theta)\}_{n\in\mathbb{Z}^+}$ and $\{h_n(\theta)\}_{n\in\mathbb{Z}^+}$ each constitute an infinite, countable set of independent, smooth functions that satisfy the above requirements, then we can almost uniquely fix the solution to be 
\begin{equation}
    f_{2n-1}(\theta) = -\sinh((2n-1)\theta) \;,\qquad g_{2n-1}(\theta) = -\theta\sinh((2n-1)\theta) \;,\qquad h_n(\theta) = \cosh(n\theta) \;,
\label{eq:f_g_h_solution}
\end{equation}
and
\begin{equation}
    f_{2n}(\theta) = 0\;,\qquad g_{2n}(\theta) = 0\;.
\label{eq:f_g_even_vanish}
\end{equation}
With ``almost uniquely'' we mean to say that -- since the equations (\ref{eq:eqf})-(\ref{eq:eqgh2}) are homogeneous -- we still have the freedom to rescale each of the functions $f_{2n-1}(\theta)$, $g_{2n-1}(\theta)$ and $h_n(\theta)$ by an arbitrary constant (in $\theta$) factor. For $f_{2n-1}(\theta)$ and $g_{2n-1}(\theta)$ this factor has to be the same, due to the first equation in \eqref{eq:eqgh}, and it can be reabsorbed into the coefficients $g^{(n)}$. However for the functions $h_n(\theta)$, it is indeed an arbitrary choice. We can even choose it to vanish for any $n$. This fact is not surprising: it is the usual indeterminacy of the $\cosh$ terms that was already remarked in previous works \cite{longpaper,Castro-Alvaredo2024compl,MMF}. Here, the choice we make (\ref{eq:f_g_h_solution}),  (\ref{eq:f_g_even_vanish}) is guided by the requirement that the functions \eqref{eq:delta_rho_ansatz} agree, for large $\vert\theta\vert$, with (\ref{eq:delta_series}), (\ref{eq:rho_series}) when stripped of their diverging behavior. So, for example, we ask $f_{2n-1}(\theta) \underset{\vert\theta\vert\to\infty}{\sim} {\rm sign}(\theta)e^{-(2n-1)\vert\theta\vert} + \textrm{``diverging terms''}$ and we see that $f_{2n-1}(\theta) = -\sinh((2n-1)\theta)$ fits the bill. Notice that this also means that
\begin{equation}
    g^{(2n)} = 0\;.
\end{equation}
Thus, we have found that the following functions (we now add an index $\bal$ to include the parameter dependence)
\begin{equation}
    \begin{split}
        \delta_{\bal}(\theta) & = - \sum_{n=1}^\infty \alpha_{2n-1} \sinh((2n-1)\theta)\;, \\
        \rho_{\bal}(\theta) & = \frac{i\pi - \theta}{2 \pi} \delta(\theta) + \sum_{n=1}^\infty \alpha_{2n-1} \left[ \frac{\cosh((2n-1)\theta)}{4\pi (2n-1)} - \frac{1}{\pi} \sum_{\genfrac{}{}{0pt}{}{m=1}{m\neq 2n-1}}^\infty \frac{m \cosh(m\theta)}{(2n-1)^2 - m^2} \right]\;, \\
        \alpha_{2n-1} &= \frac{1}{2\pi}\frac{g^{(2n-1)}}{2n-1}\;,
    \end{split}
\label{eq:delta_rho_TTbar_sol}
\end{equation}
constitute a valid $\delta(\theta)$-$\rho(\theta)$ pair. Notice that the form of the phase shift is exactly that of the $\TTb$ deformation \eqref{sum}. However, the MFF is different from that found in \cite{Castro-Alvaredo2024compl,longpaper}, in the sense that the coefficients of the $\cosh(n\theta)$ terms are not arbitrary, but related to the same $\alpha_n$ appearing in $\delta(\theta)$. This form complies instead with the expression for the sinh-Gordon MFF derived in \cite{MMF}. In fact, \eqref{eq:f_g_h_solution} is not the only possible set of smooth functions we could choose. Setting $h_n(\theta) = 0$ for all indices $n\in\mathbb{Z}$ is also an alternative choice that satisfies all the conditions. This choice reproduces the results of \cite{Castro-Alvaredo2024compl,longpaper}, when all free parameters are set to zero. We argue that the MFF in \eqref{eq:delta_rho_TTbar_sol} is a more natural choice, for at least two reasons. First of all, by construction, it reproduces the MFFs obtained from the integral representation for UV complete IQFTs.
%As an example, we can look at the sinh-Gordon theory, whose phase shift is
%
%\begin{equation}
%    \delta_{\rm shG}(\theta) = \frac{1}{i} \log\left[-\frac{\sinh\theta - i \cos \frac{\pi b}{2}}{\sinh\theta + i \cos \frac{\pi b}{2}}\right] = -4 \sum_{n = 1}^\infty (-1)^n \frac{\cos\frac{(2n-1)\pi b}{2}}{2n-1} \sinh((2n-1)\theta)\;,
%\end{equation}
%
%so that
%
%\begin{equation}
%    \alpha_{2n-1} = 4 (-1)^n \frac{\cos\frac{(2n-1)\pi b}{2}}{2n-1}\;.
%\end{equation}
%
%Then the logarithm of the MFF will be\footnote{Notice that we have to add the ``Ising term'' $\log\left[-i\sinh\frac{\theta}{2}\right]$, since $S(0) = \epsilon = -1$.}
%
%\begin{equation}
%    \rho_{\rm shG}(\theta) = \log\left[-i\sinh\frac{\theta}{2}\right] - \frac{\theta - i \pi}{2\pi}\delta_{\rm shG}(\theta)
%\end{equation}
Secondly, as a function of $\eta = \frac{\theta-i\pi}{\pi}$, the logarithm of the MFF in \eqref{eq:delta_rho_TTbar_sol} has more desirable asymptotic properties than most of the possible solutions found in \cite{Castro-Alvaredo2024compl,longpaper}. Let us look at a simple case to illustrate this fact.

\subsection{The $\TTb$-Deformed Ising Field Theory}

\begin{figure}%
    \centering
    \subfloat[\centering MFF for $\TTb$-perturbed massive free fermion with the inclusion of the $\cosh$ terms, with $\alpha = 2\pi$.]{{\includegraphics[width=.8\textwidth]{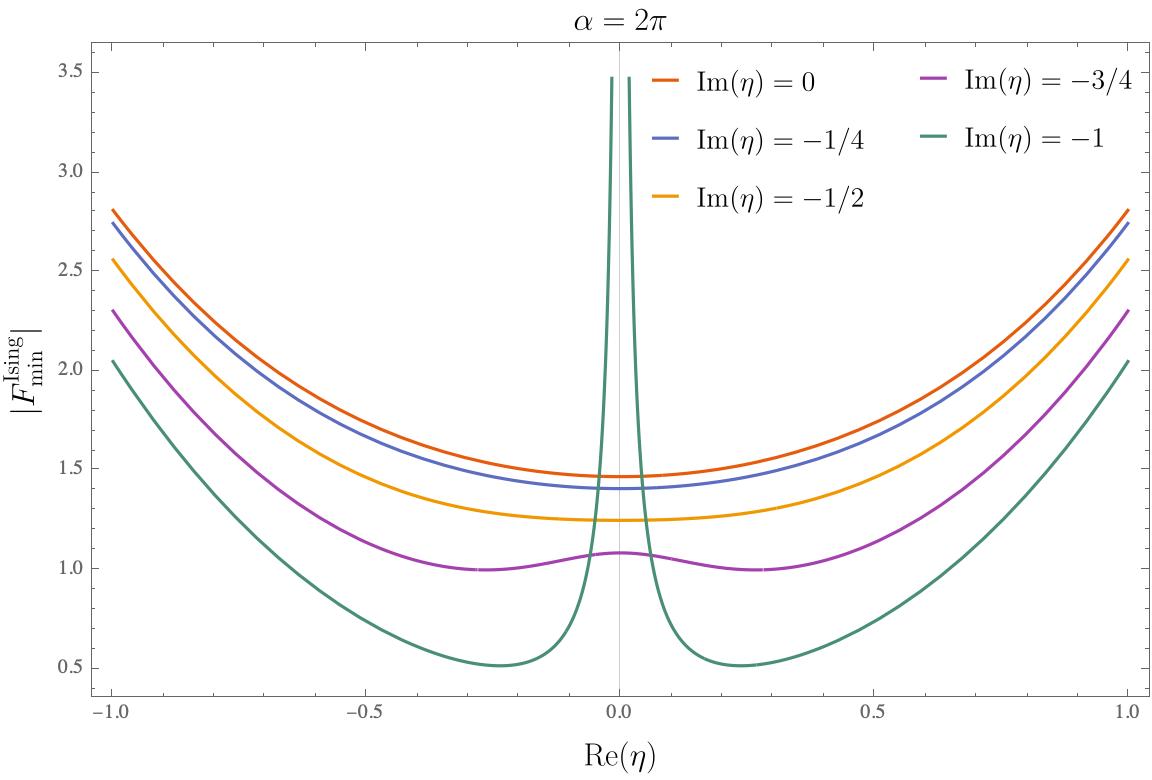} }}%
    \\[.5cm]
    \subfloat[\centering MFF for $\TTb$-perturbed massive free fermion with the inclusion of the $\cosh$ terms, with $\alpha = -2\pi$.]{{\includegraphics[width=.8\textwidth]{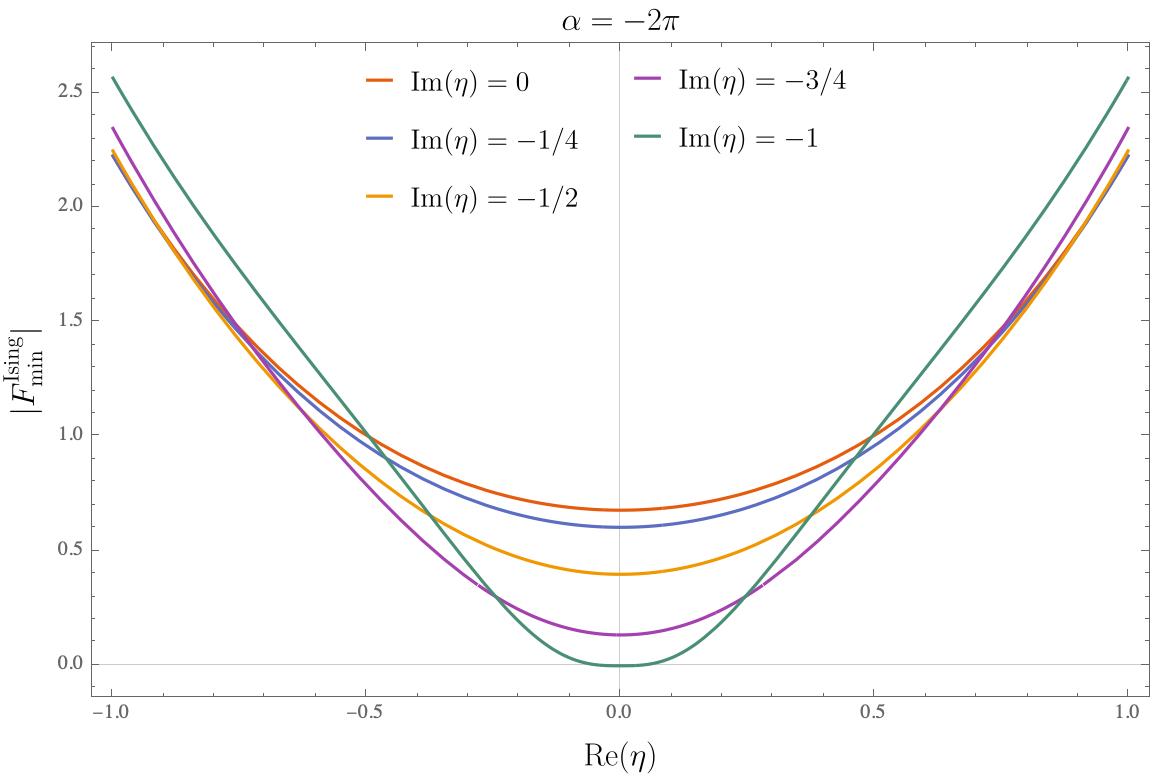} }}%
    \caption{Comparison of MFFs with the inclusion of $\cosh$ terms, for $\alpha$ positive and negative, on different lines ${\rm Im}\eta = 0,-\frac{1}{4},-\frac{1}{2},-\frac{3}{4},-1$.}%
    \label{fig:mff_on_various_lines}%
\end{figure}

\begin{figure}%
    \centering
    \subfloat[\centering MFF for $\TTb$-perturbed massive free fermion with the inclusion of the $\cosh$ terms, as a function of $\eta\in\mathbb{R}$]{{\includegraphics[width=.8\textwidth]{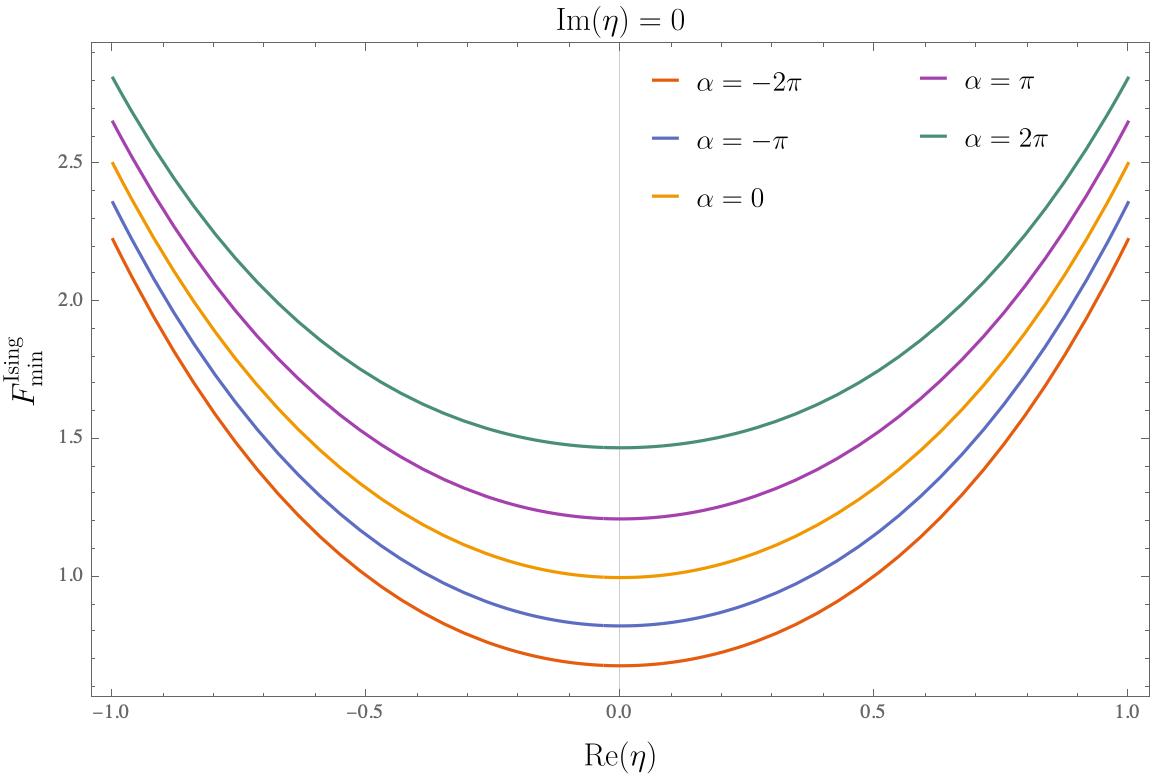} }}%
    \\[.5cm]
    \subfloat[\centering MFF for $\TTb$-perturbed massive free fermion \underline{without} the inclusion of the $\cosh$ terms (see \cite{Castro-Alvaredo2024compl,longpaper}), as a function of $\eta\in\mathbb{R}$]{{\includegraphics[width=.8\textwidth]{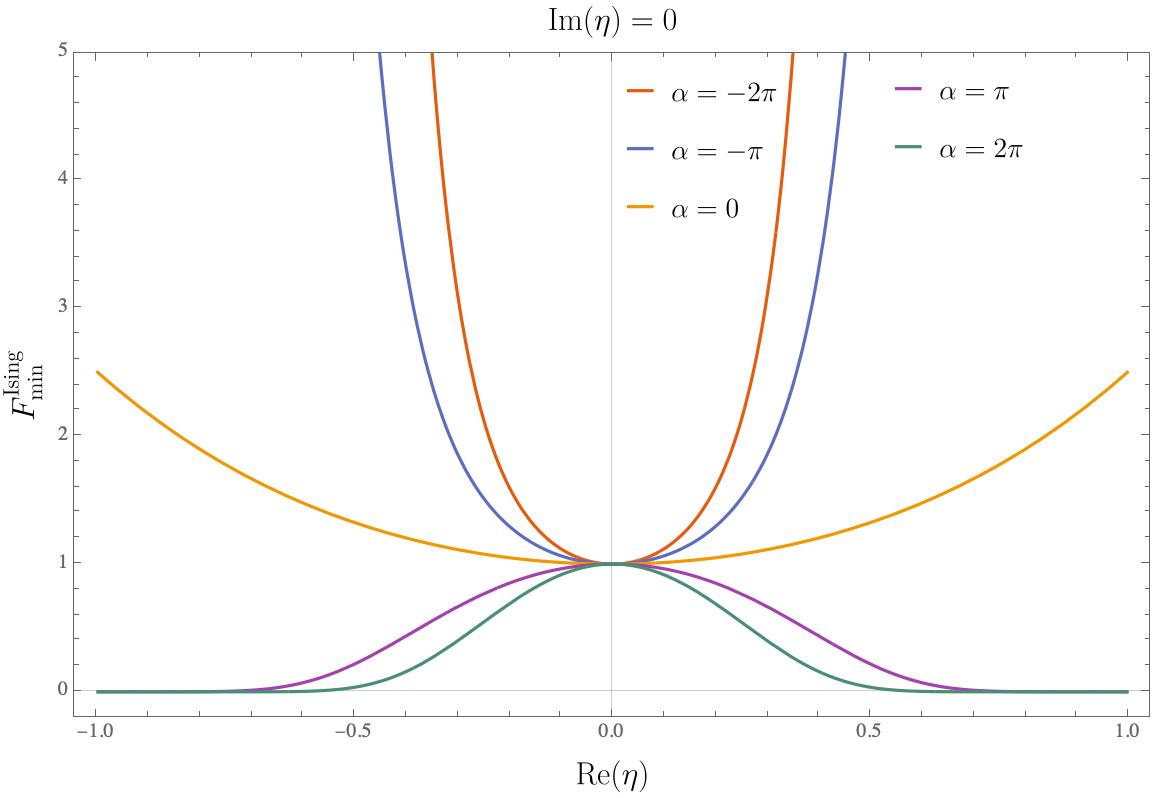} }}%
    \caption{Comparison of MFFs with and without the inclusion of $\cosh$ terms on the line $\eta \in \mathbb{R}$ (i.e. ${\rm Im}\theta = i\pi$).}%
    \label{fig:mff_comparison_eta}%
\end{figure}

\begin{figure}%
    \centering
    \subfloat[\centering MFF for $\TTb$-perturbed massive free fermion with the inclusion of the $\cosh$ terms, as a function of $\theta\in\mathbb{R}$]{{\includegraphics[width=.8\textwidth]{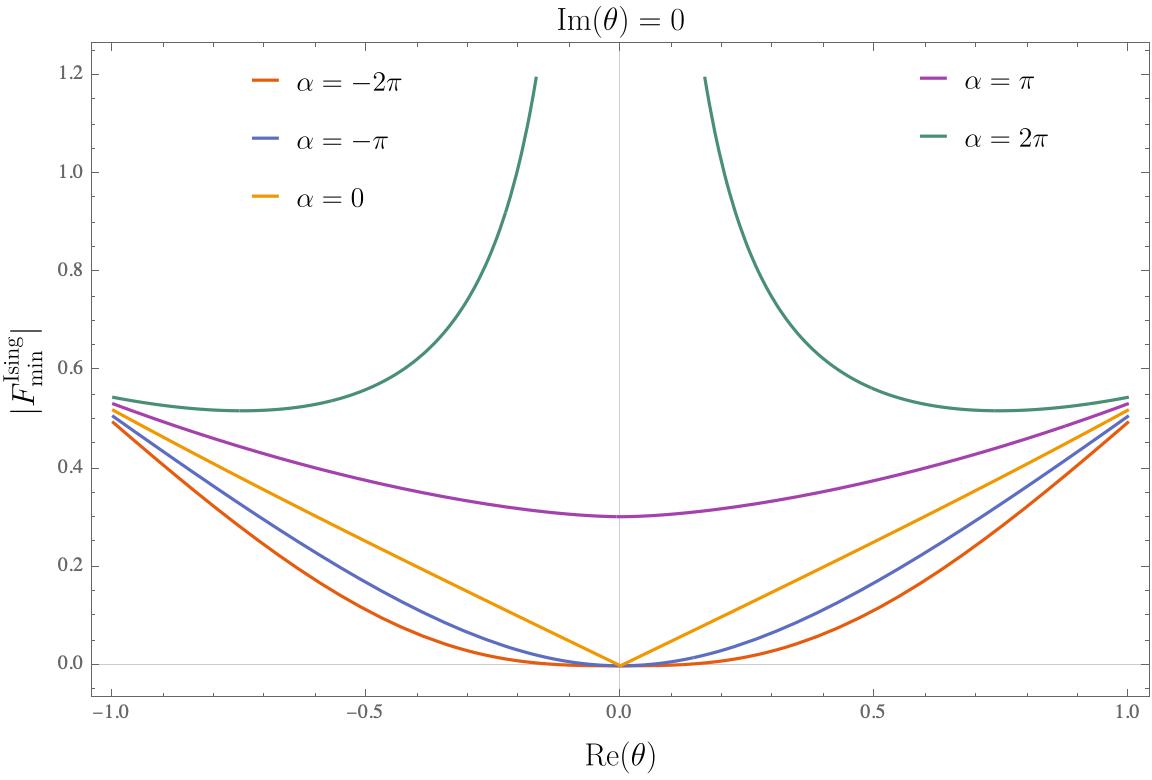} }}%
    \\[.5cm]
    \subfloat[\centering MFF for $\TTb$-perturbed massive free fermion \underline{without} the inclusion of the $\cosh$ terms (see \cite{Castro-Alvaredo2024compl,longpaper}), as a function of $\theta\in\mathbb{R}$]{{\includegraphics[width=.8\textwidth]{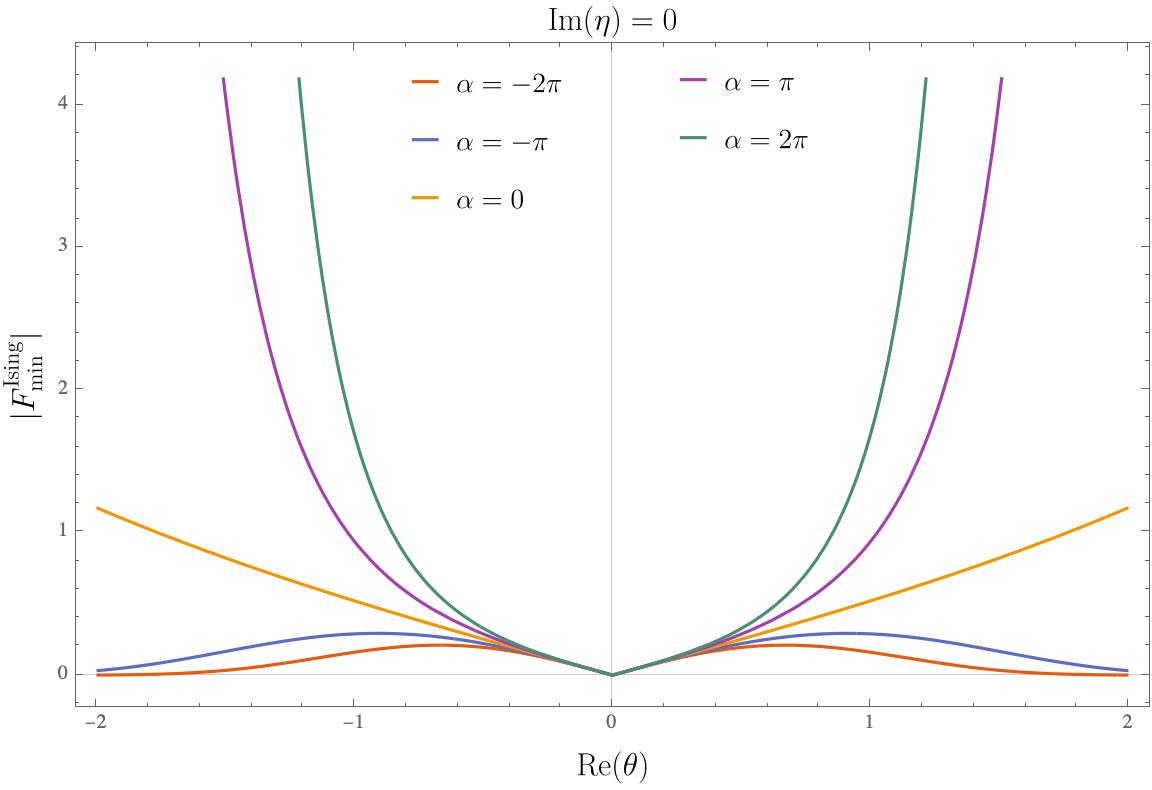} }}%
    \caption{Comparison of MFFs with and without the inclusion of $\cosh$ terms on the line $\theta \in \mathbb{R}$.}%
    \label{fig:mff_comparison_theta}%
\end{figure}

\begin{figure}
    \centering
    \includegraphics[width=0.8\linewidth]{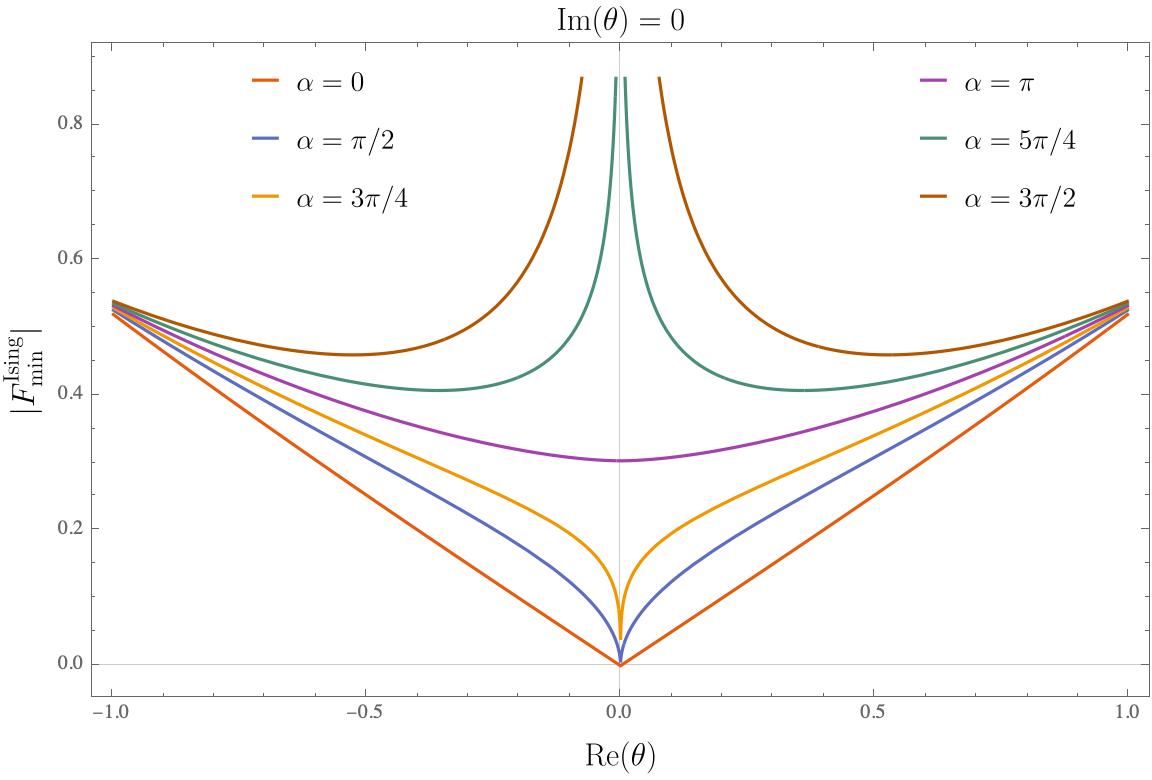}
    \caption{Plots of the MFF obtained from \eqref{eq:TTb_Ising_mff} as a function of $\theta\in\mathbb{R}$ for various values of $\alpha >0$.}
    \label{fig:mff_theta_various_alpha}
\end{figure}

Let us take the simplest possible factorized scattering theory,  a massive free fermion, with $S$-matrix $S(\theta) = -1$. Let us perform a basic $\TTb$ deformation of this theory. Then the  scattering phase becomes
\begin{equation}
    \delta_\alpha^{{\rm Ising}}(\theta) = - \alpha \sinh\theta\;,\qquad \epsilon \equiv S(0) = -1\;.
\end{equation}
From the general formula \eqref{eq:delta_rho_TTbar_sol}, we find the logarithm of the MFF to be
\begin{equation}
    \begin{split}
        \rho_\alpha^{{\rm Ising}}(\theta) &= \log\left(-i\sinh\frac{\theta}{2}\right) - \frac{i \pi - \theta}{2\pi}\alpha\sinh\theta + \frac{\alpha}{4\pi}\cosh\theta - \frac{\alpha}{\pi} \sum_{m=2}^\infty \frac{m\, \cosh(m\theta)}{1-m^2}  \\
        &= \log\left(-i\sinh\frac{\theta}{2}\right) - \frac{i \pi - \theta}{2\pi}\alpha\sinh\theta  - \frac{\alpha}{2\pi} - \frac{\alpha}{2\pi}\cosh\theta \log(-4\sinh^2\frac{\theta}{2}) \\
        & = \log\left(\cosh\frac{\pi\eta}{2}\right) - \frac{\alpha}{2\pi}\left[1  + \pi\eta  \sinh(\pi\eta) - \cosh(\pi\eta) \log\left(4 \cosh^2\frac{\pi\eta}{2}\right)\right]\;,
    \end{split}
\label{eq:TTb_Ising_mff}
\end{equation}
where we now reinstated the ``Ising term''\footnote{Notice that, thanks to this term, the minimal form factor $F_{{\rm min}}^{\rm Ising}(\theta;\alpha) = \exp\rho_{\alpha}^{\rm Ising}$ diverges as ${\rm Re}\theta\to\pm\infty$. This is a characteristic behaviour of the massive free fermion. In non-free theories, this diverging behaviour will be tempered by the presence of interactions.} $\log\left(-i\sinh\frac{\theta}{2}\right)$, since $\epsilon = -1$. 
We introduced the variable $\eta = \frac{\theta-i\pi}{\pi}$. This function decays nicely for $\vert\theta\vert\to\infty$ in the whole strip $0\leq {\rm Im}\,\theta \leq 2\pi$, for both $\alpha$ positive and negative, as can be seen in Figure \ref{fig:mff_on_various_lines}. Figures \ref{fig:mff_comparison_eta} and \ref{fig:mff_comparison_theta} show a comparison of the MFF obtained from \eqref{eq:TTb_Ising_mff} with that derived in \cite{Castro-Alvaredo2024compl,longpaper}, for, respectively, ${\rm Im}\theta = i\pi$ and ${\rm Im}\theta = 0$. We notice how the MFF we obtained here preserves the large-rapidity behavior of the unperturbed ($\alpha = 0$) theory, while the one obtained in \cite{Castro-Alvaredo2024compl,longpaper} agrees with the $\alpha = 0$ situation only in the vicinity of $\theta = 0$.
This difference in the behaviour of the deformation of the MFF is more profound than appears at a cursory glance. The expression found in \cite{Castro-Alvaredo2024compl,longpaper} cannot be considered a small deformation of the massive free fermion MFF! This was already remarked in \cite{Castro-Alvaredo2024compl,longpaper}, where it was observed how the deformation leads to a radical difference in the correlation functions compared to the undeformed theory. In contrast, the expression \eqref{eq:TTb_Ising_mff}, can be rightfully considered a well-behaved $1$-parameter ($\alpha$) deformation of the massive free fermion MFF. This is immediately clear from the plot of $e^{\rho_\alpha^{\,{\rm Ising}}(\theta)}$ in Figure \ref{fig:mff_comparison_eta}. From the plot in Figure \ref{fig:mff_comparison_theta}, we see that something peculiar happens for $\alpha>0$: the minimal form factor seems to change its $\theta\sim0$ behaviour when $\alpha$ crosses the value $\alpha_c=\pi$. In fact, a more careful analysis confirms this observation. Let us consider the behaviour of the MFF $F_{\rm min}^{\rm Ising}(\theta;\alpha):=e^{\rho_\alpha^{\,{\rm Ising}}(\theta)}$ in the vicinity of $\eta = 0$ and $\theta = 0$. 
\begin{equation}
    \begin{split}
        F_{\rm min}^{\rm Ising}(\theta;\alpha) & \underset{\eta\to0}{\sim} e^{ -\frac{\alpha(1 - 2\log2)}{2\pi}} + O(\eta^2)\;, \\
        F_{\rm min}^{\rm Ising}(\theta;\alpha) & \underset{\theta\to0}{\sim} \frac{1}{2}(-\theta^2)^{\frac{1}{2}-\frac{\alpha}{2\pi}}e^{-\frac{\alpha}{2\pi}} \left[1 + \frac{\alpha}{2} (-\theta^2)^{\frac{1}{2}} + O(\theta^2) \right]\;.
    \end{split}
\end{equation}
We immediately see that the behaviour at $\eta \to 0$ is qualitatively the same for any value of $\alpha$: the MFF is a finite, $\alpha$-dependent constant. On the other hand, we see that at $\theta = 0$, the behaviour of the MFF radically depends on the value of $\alpha$: if $\alpha < \pi$, it vanishes (as happens for the undeformed case), while for $\alpha > \pi$ it diverges (reaching a constant in the limit case $\alpha = \alpha_c = \pi$). The situation is portrayed in Figure \ref{fig:mff_theta_various_alpha}.
Further discussion of this behaviour in the context of correlation functions is given in Section \ref{correlation}.

Notice that the critical value $\alpha_c=\pi$ is theory-dependent. In this case, it results from the fact that the MFF of the Ising field theory is $-i\sinh\frac{\theta}{2}$, as we see for instance in equation (\ref{eq:TTb_Ising_mff}). The general statement is that if the leading asymptotic behaviour of the MFF of the unperturbed theory is:
\beq 
F_{\rm min}(\theta)\underset{\theta\to0}{\sim}  \theta^\kappa\,,
\eeq
for some constant $\kappa \in \mathbb{R}$,
then, the MFF of the $\TTb$-perturbed theory will have the leading asymptotics
\beq 
F_{\rm min}(\theta;\alpha) \underset{\theta\to0}{\sim}  \theta^{\kappa-\frac{\alpha}{\pi}}\,,
\label{59}
\eeq
so that there is a critical value $\alpha_c=\kappa \pi$ above which the MFF develops a pole of order $\frac{\alpha}{\pi}-\kappa$ at $\theta=0$. In the Ising field theory we simply have $\kappa=1$.

\subsection{The $\TTb_{2n-1}$-Deformed Ising Field Theory}
\label{generalcase}
Let us now consider a generalisation of the computation in the previous subsection, to the case when the perturbation is a single spin-$2n-1$ irrelevant perturbation. That is:
\begin{equation}
    \delta_{\alpha_{2n-1}}^{{\rm Ising}}(\theta) = - \alpha_{2n-1} \sinh((2n-1)\theta)\;,\qquad \epsilon \equiv S(0) = -1\;.
\end{equation}
From the general formula \eqref{eq:delta_rho_TTbar_sol}, we find the logarithm of the MFF to be
\beqa
        \rho_{\alpha_{2n-1}}^{{\rm Ising}}(\theta) &=& \log\left(-i\sinh\frac{\theta}{2}\right)  + \frac{\theta - i \pi}{2\pi}\alpha_{2n-1}\sinh((2n-1)\theta) +\frac{\alpha_{2n-1}}{4\pi(2n-1)}\cosh((2n-1)\theta)\nonumber\\ &&  - \frac{\alpha_{2n-1}}{\pi} \sum_{\genfrac{}{}{0pt}{}{m=1}{m\neq 2n-1}}^\infty \frac{m\, \cosh(m\theta)}{(2n-1)^2-m^2}\,,
\label{eq:TTb_Ising_mff2}
\eeqa
The sum above can be computed as follows:
\beq 
\sum_{\genfrac{}{}{0pt}{}{m=1}{m\neq s}}^\infty \frac{m\, \cosh(m\theta)}{s^2-m^2}= \sum_{m=1}^{s-1} \frac{m\, \cosh(m\theta)}{s^2-m^2}+ \sum_{m=s+1}^{\infty} \frac{m\, \cosh(m\theta)}{s^2-m^2}\,,
\eeq 
where the infinite sum is given by
\beqa 
\sum_{m=s+1}^{\infty} \frac{m\, \cosh(m\theta)}{s^2-m^2} =  \frac{1}{2s} + \frac{1}{4s} \cosh(s\theta) + \sum_{m=1}^{s-1} \frac{s\, \cosh(m\theta)}{s^2-m^2} + \frac{1}{2} \cosh(s\theta) \log\left(-4\sinh^2\frac{\theta}{2}\right)\,,
\eeqa 
so that 
\beq 
\sum_{\genfrac{}{}{0pt}{}{m=1}{m\neq s}}^\infty \frac{m}{s^2-m^2}\cosh(m\theta)= \frac{1}{2s} + \frac{1}{4s} \cosh(s\theta)+\sum_{m=1}^{s-1} \frac{\cosh(m\theta)}{s-m}+\frac{1}{2} \cosh(s\theta) \log\left(-4\sinh^2\frac{\theta}{2}\right)\,.
\label{othersum}
\eeq 
Putting everything together we have that the logarithm of the minimal form factor is
\beqa
        \rho_{\alpha_{2n-1}}^{{\rm Ising}}(\theta) &=& \log\left(-i\sinh\frac{\theta}{2}\right)  + \frac{\theta - i \pi}{2\pi}\alpha_{2n-1}\sinh((2n-1)\theta) -\frac{\alpha_{2n-1}}{2\pi (2n-1)}\nonumber\\ &&   - \frac{\alpha_{2n-1}}{\pi}\sum_{m=1}^{2n-2} \frac{\cosh(m\theta)}{2n-1-m} - \frac{\alpha_{2n-1}}{2\pi} \cosh((2n-1)\theta)\log\left(-4\sinh^2\frac{\theta}{2}\right)\,.
\label{eq:TTb_Ising_mff3}
\eeqa
For the purpose of studying the asymptotic properties of the minimal form factor, it is useful to rewrite the function above in terms of the variable $\eta$, as in the previous subsection
\beqa
        \rho_{\alpha_{2n-1}}^{{\rm Ising}}(\theta) &=& \log\left(\cosh\frac{\pi\eta}{2}\right)  - \frac{\alpha_{2n-1}}{2\pi}\left[\frac{1}{2n-1}+\pi \eta\sinh((2n-1)\pi\eta) \right]\nonumber\\ &-&  \frac{\alpha_{2n-1}}{\pi}\sum_{m=1}^{2n-2} \frac{(-1)^m \cosh(m \pi \eta)}{2n-1-m} +\frac{\alpha_{2n-1}}{2\pi} \cosh((2n-1)\pi \eta)\log\left(4\cosh^2\frac{\pi\eta}{2}\right)\,.
\label{eq:TTb_Ising_mffeta}
\eeqa
We then find that 
\begin{equation}
    \begin{split}
        F_{\rm min}^{\rm Ising}(\theta;\alpha_{2n-1}) & \underset{\eta\to0}{\sim} e^{-\frac{\alpha_{2n-1}(1 - (2n-1) 2 (\log2+c_{2n-1}))}{2\pi (2n-1)}} + O(\eta^2)\;, \\
        F_{\rm min}^{\rm Ising}(\theta;\alpha_{2n-1}) & \underset{\theta\to0}{\sim} \theta^{1-\frac{\alpha_{2n-1}}{\pi}}e^{-\frac{\alpha_{2n-1} \left(1+ 2(2n-1) \hat{c}_{2n-1})\right)}{2\pi(2n-1)}} \left[1-\frac{i\alpha_{2n-1} \theta}{2} + O(\theta^2) \right]\;,
    \end{split}
\end{equation}
where
\beq 
c_{2n-1}=\sum_{m=1}^{2n-2}\frac{(-1)^{m+1}}{m}\,\qquad \mathrm{and} \qquad \hat{c}_{2n-1}=\sum_{m=1}^{2n-2}\frac{1}{m}\,,
\label{cn}
\eeq 
which are very similar behaviours as found for the $n=1$ case in (\ref{59}). In particular, we find again that there is a stark change in the $\theta \rightarrow 0$ asymptotics depending on whether $\alpha_{2n-1}$ is greater or smaller than $\pi$. It is also easy to see that in the presence of several perturbations the critical value will correspond to solutions to $\pi \kappa-\sum_{s\in \mathcal{S}} \alpha_s=0$ where $\kappa $ is the value in (\ref{59}).

%{\SN I do not think it makes sense to insert additional plots for the cases $n>1$: they look pretty much identical to the $n=1$ ones, with some sharper divergences. If you want I can produce them without too much hassle, but I think they are superfluous.}{\OCA Agree!}
%{\OCA My asymptotics for large $\theta$ is different from Stephano's for n=1... He gets $\theta^{-1-\alpha/\pi}$ and I am getting $\theta^{1-\alpha/\pi}$ which makes the value $\pi$ the critical one for me... I think this is to do with the sign of the $\log(\sinh\frac{\theta}{2})$ term as discussed earlier.}
%{\OCA Fabio wrote a discussion talking about this idea of a particle length. Perhaps another way (or maybe the same way) is to think that with this form factor the big difference is not between $\alpha>0$ and $\alpha<0$ but between $\alpha>\pi$ and $\alpha<\pi$ (which would include $\alpha$ negative). I still need to write the discussion on correlation functions where we will probably see this playing  a role again.}

\section{Alternative Construction from the Karowski-Weiss Approach}
\label{section3}

\subsection{The Original work}
\begin{figure}[h]
    \centering
\includegraphics[width=0.7\linewidth]{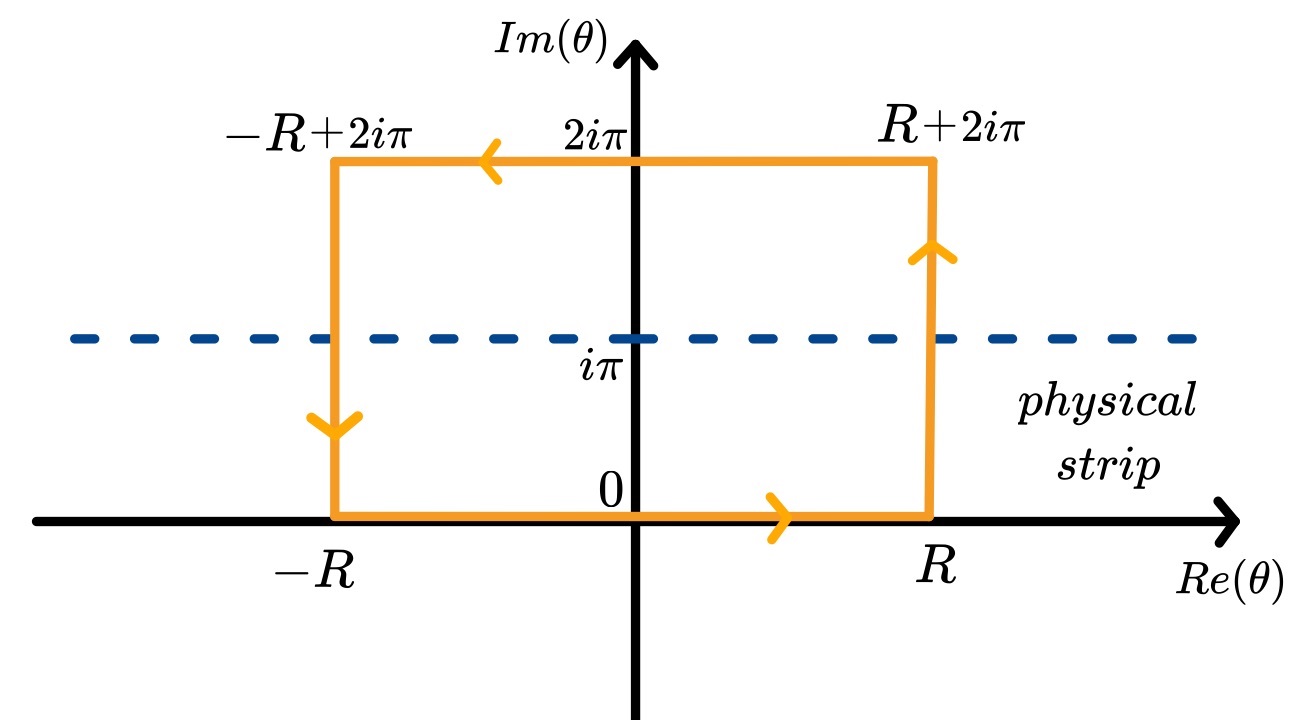}
    \caption{The contour $C$ in equation (\ref{KW1}).}
    \label{KWcontour}
\end{figure}

Let us go back to our original equations (\ref{FFE}). We consider once more the problem of finding general solutions with desirable asymptotic properties to these equations for  $\TTb$-perturbed theories. An alternative way to approach this problem is suggested by the original construction of Karowski and Weisz which leads also to an integral representation of the MFFs, albeit written in a slightly different way. In this subsection we summarise their original work. 
We start by writing
\beq
\frac{\p}{\p \theta}\log{F_{\rm min}(\theta)}= \frac{1}{8 \pi i} \int_C \frac{\log{F_{\rm min}(z)}}{\sinh^2{\frac{1}{2}(z-\theta)}} dz \,.
\label{KW1}
\eeq
Here $C$ is a contour enclosing the strip $0 \leq \textrm{Im}(\theta) \leq 2 \pi$ as shown in Fig.~\ref{KWcontour}. 

The denominator $\sinh^2{\frac{1}{2}(z-\theta)}$ has double poles at $z= \theta+ 2 \pi i \, n$ ($n \in \mathbb{N}$). For $\theta>0$ only the pole at $z=\theta$ falls within the contour. Then, by Cauchy's residue theorem, we obtain the derivative on the l.h.s. Let us examine this construction better. Because of the equations \eqref{FFE}, we have certain asymptotic properties the form factor must satisfy. If the $S$-matrix tends to a constant in the high energy limit $\theta \rightarrow \infty$, as is expected in UV-complete theories, then we must have  $F_{\rm min}(\theta)= O (\exp{\exp{|\theta|}})$ for ${\rm Re}(\theta) \to \infty$.  This is very important when computing the integral along the contour above. We can divide the contour $C$ in  in four pieces: $C= C_1+C_2+C_3+C_4$. Given $z=x+i y$ we have:
\begin{itemize}
    \item two horizontal segments: $C_1= \left\{ x: -R \leq x \leq R    \right\}$ and $C_3= \left\{ x+ 2 i \pi: -R \leq x \leq R    \right\}$;
    \item two vertical segments: $C_2= \left\{R+ i y: 0 \leq y \leq 2 \pi    \right\}$ and $C_4= \left\{-R+ i y: 0 \leq y \leq 2  \pi    \right\}$;
\end{itemize}
with $C$ oriented counterclockwise. So, our integral can be written also as
\begin{equation}
    \int_C dz \,w(z)= \int_{-R}^{R}dx\, w(x) + \int_{R}^{-R}dx \, w(x+ 2 \pi i)  + i \int_{0}^{2 \pi}dy \, w(R+i y)+ i\int_{2 \pi}^{0}dy \, w(-R+i y)\,,
\end{equation}
with 
\beq 
w(z)=\frac{\log{F_{\rm min}(z)}}{8\pi i \sinh^2{\frac{1}{2}(z-\theta)}}\,.
\eeq 
The idea is that, if $w(z)$ decays fast enough for large $R$, we can discard the vertical contributions. That is by the estimation lemma we can write
\begin{equation}
    \left|\int_{C_2} dz \, w(z) \;\right|  \leq \int_0^{2 \pi} dy \, |w(R+ i y)| \underset{R\rightarrow \infty}{\rightarrow}  0
\end{equation}
and similarly for the integral along $C_4$. For our function $w(z)$ we have that  
\begin{equation}
    \left| \frac{\log{F_{\rm min}(R+i y)}}{\sinh^2{\frac{1}{2}(R-\theta+i y)}}\right| \sim \frac{O(e^{|R|})}{e^{|R|}} \underset{R\rightarrow \infty}{\rightarrow}  0\,,
\end{equation}
 because the MFF has the asymptotic behaviour described above. Therefore, the integral is given by the sum of the horizontal contributions. These contributions are related to each other by the form factor equations (\ref{FFE}) giving 
\beqa
\frac{1}{8 \pi i} \int_C \frac{\log{F_{\rm min}(z)}}{\sinh^2{\frac{1}{2}(z-\theta)}} dz &=& 
 \frac{1}{8 \pi i} \int_{-\infty}^{\infty} \frac{dx}{\sinh^2{\frac{1}{2}(x-\theta)}} \log{\frac{F_{\rm min}(x)}{F_{\rm min}(x+2 \pi i)}} \nonumber \\
 &=& \frac{1}{8 \pi i} \int_{- \infty}^{\infty} \frac{\log{S(x)}}{\sinh^2{\frac{1}{2}(x-\theta)}} dx\,,
 \label{KW2}
\eeqa
which therefore provides a representation of the derivative of the MFF in terms of the scattering matrix.  
If the S-matrix admits a Fourier series such as \eqref{Smatrixfourier}, then a representation of the type \eqref{KWFourierMFF} follows immediately.

\subsection{Challenges Posed by $\TTb$-Deformations}
Consider now an $S$-matrix of the type \eqref{Smatrix} with a CDD factor of the $\TTb$ type as in (\ref{sum}). Let $\mathcal{D}_{\bal}(\theta)$ be the deformation of the corresponding MFF\footnote{In (\ref{minimal}) and (\ref{betas}) we wrote the most general solution to these equations, namely  $\mathcal{D}_{\bal}(\theta)=\varphi_{\bal}(\theta)C_{\bel}(\theta)$.}, as defined by the equations (\ref{DMM}) whose general solution we wrote in (\ref{minimal}). 
The main property of the CDD factor $\Phi_{\bal}(\theta)$ is that, at large rapidities it does not tend to a constant.   Therefore in general we have $\mathcal{D}_{\bal}(\theta) \sim \exp{(\exp{s^* |\theta|})}$ for ${\rm Re}(\theta)\to \infty$ and $s^*$ the largest spin in the sum (\ref{sum}). This means that if we follow again the Karowski-Weisz construction we find the following asymptotics of $w(R+i y)$

\begin{equation}
    \left| \frac{\log{\mathcal{D}_{\bal}(R+i y)}}{\sinh^2{\frac{1}{2}(R-\theta+i y)}}\right| \sim \frac{e^{s^*|R|}}{e^{|R|}} \underset{R\rightarrow \infty}{\rightarrow}  e^{(s^*-1)|R|}\,.
\end{equation}
So we can no longer argue that the vertical contributions to the integral are vanishing. 
We can think of two ways, possibly related to each other, of solving this problem: the first consists in finding an appropriate choice of contour such that the exponential growth of the function on the real axis can be mitigated.
The second comes from the freedom given by the equations in \eqref{RHproblem} regarding the choice of the MFF. As noticed in the introduction, we can always multiply a given solution by some function $C_{\bel}(\theta)$ given by (\ref{betas}). Let us call this modified solution $\tilde{\mathcal{D}}_{\bal}(\theta)$.
An appropriate choice of this function, which means a choice of the parameters $\bel$, will ensure that 
\begin{equation}
    \left| \frac{\log{\tilde{\mathcal{D}}_{\bal}(R+i y)}}{\sinh^2{\frac{1}{2}(R-\theta+i y)}}\right| \sim \frac{O(e^{|R|})}{e^{|R|}} \underset{R\rightarrow \infty}{\rightarrow}  0\,.
\end{equation}
The additional parameters $\bel$, if chosen wisely, can enforce the double-exponential asymptotics of the MFF \cite{MMF}. The idea is therefore to regularise the usual integral representation by imposing this asymptotics. 
\medskip

We end this section with an observation. In general, given an integrand $I(z,\theta)= \frac{\log{S(z)}}{\sinh^2{\frac{1}{2}(z-\theta)}}$ over a contour such that $\log(S(z))$ is analytic everywhere inside and on the contour and $z_n$ is a pole of the denominator, we have that the residue at the pole is given by
\beq
\Res\left( I(z,\theta),z_n\right)= 4 \p_z \left[ \log{S(z)} \right]_{z=z_n}\,.
\eeq
As observed previously, the poles are exactly at $z_n= \theta+ 2 \pi i n$, for $n=0,1,2,\dots$. 
If we consider again the contour $C$ above, including the pole at $z_0=\theta$ we get
\beq
\oint_{C} dz\, I(z,\theta)= 4 \p_z \left[\log{S(z)}\right]_{z_0}\,.
\label{residueMFF}
\eeq
 Instead of deforming the contour we will see in the next section that imposing the asymptotics above is sufficient to find a closed solution that fixes the MFF completely. Not coincidentally
the term that regularises the integral representation is proportional to contributions of the form \eqref{residueMFF}.

\subsection{A Solution}
\label{solution}

The main problem in defining a complete MFF for a generalized $\TTb$-deformation comes from the distinct properties of the CDD-factors. We cannot expand them in terms of a Fourier series, therefore we need to start from an integral representation of the type \eqref{KW2}. In practice, it is more convenient to start from an equivalent representation which involves $\log S(z)$ explicitly. This kind of representation appears in \cite{niedermaier1995freefieldrealizationform}:
\beq
\rho(\theta) =   \frac{1}{i \pi} \cosh^2{\frac{\theta}{2}} \int_0^{\infty} dt \frac{ \tanh\frac{t}{2} \ln{S(t)}}{ \cosh{t} - \cosh{\theta} }   \,, \qquad \textrm{for}\; 0 \leq {\rm Im}(\theta) \leq 2 \pi \,.
\label{NiedermeierMFF}
\eeq
The integrand reproduces exactly the same pole structure of \eqref{KW1} but it can also be shown to be equivalent to \eqref{KWFourierMFF} once we invert the Fourier transform. Let us consider $\Phi_{\bal}(\theta)$ as defined in (\ref{sum}). We have 
\beqa
\log{\mathcal{D}_{\bal}(\theta)}= \frac{1 }{i \pi} \cosh^2{\frac{\theta}{2}} \int_0^{\infty} dt \frac{ \tanh \frac{t}{2} \ln{\Phi_{\bal}(t)}}{ \cosh{t} - \cosh{\theta} }= -\frac{1}{ \pi} \cosh^2{\frac{\theta}{2}} \sum_{s \in \mathcal{S}} \int_0^{\infty} dt \frac{\alpha_s \tanh\frac{t}{2}\sinh{st}}{ \cosh{t} - \cosh{\theta}}\,.
\label{62}
\eeqa
This integral is not convergent because of the asymptotics of the $\sinh(s t)$ function, however it can be easily regularised by replacing $\sinh(s t)$ by $-e^{-s t}$. This amounts to the regularisation prescription
\beq
\log{\mathcal{D}_{\bal}(\theta)}=  \frac{1}{ i \pi}  \cosh^2{\frac{\theta}{2}} \int_0^{\infty} dt \frac{\tanh\frac{t}{2} }{ \cosh{t} - \cosh{\theta} } \left( \ln{\Phi_{\bal}(t)}-  \p_t \ln{\hat{\Phi}_{\bal}(t)} \right)\;,
\label{niceformula}
\eeq
where 
\beq 
\ln \hat{\Phi}_{\bal}(\theta)=-i \sum_{s\in 2 \mathbb{Z}^{+} -1} \frac{\alpha_s}{s} \sinh(s\theta)\,.
\label{phihat}
\eeq 
Let us see explicitly how this prescription leads to the same formulae found in the previous section. 
For $| \textrm{Re}(\theta)|<\textrm{Re}(t)$  the integrand can be expanded as a \textit{long-wave expansion} in $\theta$  \cite{Zamolodchikov:1991MasslessFlows, niedermaier1995freefieldrealizationform} giving
\beq
\frac{\cosh^2{\frac{\theta}{2}}\tanh{\frac{t}{2}}}{\cosh{t}-\cosh{\theta}}= \sum_{m = 1}^{\infty} \left[ \cosh(m \theta)+(-1)^{m+1} \right] e^{-m t}\,.
\label{longwaveexp}
\eeq
Consider next the case of a single perturbation of spin $s$.  We get then 
\beq
\log{\mathcal{D}_{\alpha_s}(\theta)}= -\frac{\alpha_s}{ \pi} \sum_{m= 1}^{\infty} \left[ \cosh{k \theta}+(-1)^{m+1} \right] \int_0^{\infty} dt  e^{-m t} \sinh{st}\,.
\eeq
We are left with the computation of a simple integral
\beq
 \int_0^{\infty} dt  e^{-m t} \sinh{st}= \frac{s}{m^2-s^2}\,,
 \label{65}
\eeq
which is only well defined away from the simple pole at $m>s$. This means that we need to regularise the infinite sum in $k$ of \eqref{longwaveexp}
by subtracting the pole. We can do this by modifying (\ref{65}) to 
\beq
\frac{s}{m^2-s^2} -\frac{m}{m^2-s^2}= - \frac{1}{m+s}\,,
\label{66}
\eeq
which is equivalent to the prescription (\ref{niceformula}) for each term in the sum in $s$.
The regularised version reads
\beq
\log{\mathcal{D}_{\alpha_s}(\theta)}= \frac{ \alpha_s}{ \pi} \sum_{m = 1}^{\infty} \frac{\cosh(m \theta)+(-1)^{m+1}}{m+s} \,.
\label{79}
\eeq
Let us consider carefully the various contributions to this sum, taking care to compare them to the result of Subsection \ref{generalcase}. First, we have the constant
\beqa 
\sum_{m = 1}^{\infty} \frac{(-1)^{m+1}}{m+s} =- \sum_{m= s+1}^{\infty} \frac{(-1)^{m+1}}{m}= -\sum_{k = 1}^{\infty} \frac{(-1)^{m+1}}{m}+ \sum_{m = 1}^{s} \frac{(-1)^{m+1}}{m}= - \log 2 + \frac{1}{s}+ c_{s}\,,
\eeqa 
where $c_{s}$ is the constant we defined earlier (\ref{cn})
and $s$ is odd.

Then we have the sum involving $\cosh(m\theta)$. It is useful to consider the two terms in (\ref{66}) separately and exclude the problematic value $m=s$. We compute
\beq 
\sum_{\genfrac{}{}{0pt}{}{m=1}{m\neq s}}^{\infty} \frac{s\cosh(m \theta)}{m^2-s^2} = \sum_{m=1}^{s-1} \frac{s\cosh(m \theta)}{m^2-s^2}  + \sum_{m=s+1}^{\infty} \frac{s\cosh(m \theta)}{m^2-s^2}\,. 
\eeq 
The infinite sum can be computed to 
\beq 
\sum_{m=s+1}^{\infty} \frac{s\cosh(m \theta)}{m^2-s^2}=-\sum_{m=1}^{s-1} \frac{s\cosh(m \theta)}{m^2-s^2}+\frac{1}{2s}+\frac{\cosh(s\theta)}{4s}-\frac{\theta-i\pi}{2}\sinh(s\theta)\,,
\eeq 
so that 
\beq 
\sum_{\genfrac{}{}{0pt}{}{m=1}{m\neq s}}^{\infty} \frac{s\cosh(m \theta)}{m^2-s^2}=\frac{1}{2s}+\frac{\cosh(s\theta)}{4s}-\frac{\theta-i\pi}{2}\sinh(s\theta)\,.
\eeq 
The other sum in (\ref{66}) was computed in (\ref{othersum}). In addition, we have the $m=s$ term coming from the $\cosh(m\theta)$ sum in (\ref{79}) which is simply $-\frac{\cosh(s\theta)}{2s}$ and cancels with contributions from the sums above. Putting everything together with $s=2n-1$ we get
\beqa 
\log\mathcal{D}_{\alpha_{2n-1}}(\theta)&=&\frac{\theta-i\pi}{2\pi} \alpha_{2n-1} \sinh((2n-1)\theta)+\frac{\alpha_{2n-1}}{\pi}\left(- \log 2 + c_{2n-1}\right)\\
&& -\frac{\alpha_{2n-1}}{\pi}\sum_{m=1}^{2n-2} \frac{\cosh(m\theta)}{2n-1-m}-\frac{\alpha_{2n-1}}{2 \pi} \cosh((2n-1)\theta) \log\left(-4\sinh^2\frac{\theta}{2}\right)\,.\nonumber
\eeqa 
Compared to the solution (\ref{eq:TTb_Ising_mff3}) without the Ising term, we see that they are identical up to normalisation constants. These are not important in applications, since the MFF is ultimately normalised by its value at $i\pi$. Generalising to any number of $\TTb$ perturbations and employing the standard normalisation we can write the general formula (\ref{original}).
If we compare this solution to the general formula we wrote in \cite{longpaper,Castro-Alvaredo2024compl}, we have that the $\bel$-dependent part in (\ref{betas}) is now fully fixed to
\beq 
C_{\bel}(\theta)=C_{\bel}(i\pi) \prod_{s \in 2\mathbb{Z}^{+}-1}\left[-\frac{1}{2}\,e^{c_s-\sum\limits_{m=1}^{s-1}\frac{\cosh(m\theta)}{s-m}} \left(2i \sinh{\frac{\theta}{2}} \right)^{-\cosh{(s\theta)}}\right]^{\frac{\alpha_s}{\pi}}\,.
\label{newC}
\eeq 
and $\varphi_{\bal}(\theta)$ is the same function as in \eqref{betas}. This construction can be easily extended to branch point twist fields, which satisfy slightly different form factor equations \cite{entropy,ourTTb3}. We discuss this generalisation in Appendix \ref{BPTFs}.
\section{Correlation Functions and their Asymptotic Properties}
\label{correlation}
One of the main uses of the form factor program is for the computation of correlation functions of local fields in the ground state. Such correlators can be expanded in terms of integrals of the absolute value squared of the form factors. In our papers \cite{longpaper,Castro-Alvaredo2024compl} we argued that the correlation functions resulting from a MFF where the function $\mathcal{D}_{\bal}(\theta)$ is chosen as $\mathcal{D}_{\bal}(\theta)=e^{\frac{\theta-i\pi}{2\pi} \alpha \sinh\theta}$ will contain integrals whose integrand contains the absolute value of this function squared, that is, $e^{\frac{\alpha \theta}{\pi}\sinh\theta}$. This function grows extremely rapidly for $|\theta|$ large and $\alpha>0$ and decays extremely rapidly if $\alpha<0$. Therefore, there are very different physical properties in these two regimes, a feature that is (qualitatively) in line with what is expected from the TBA analysis \cite{Cavaglia:2016oda}.

Consider instead the new function (\ref{original}). For simplicity, let us take $s=1$ and $\alpha_1=\alpha$ so that we have a single $\TTb$ perturbation and a single perturbation parameter. The absolute value squared gives
\beq 
|\mathcal{D}_{\bal}(\theta)|^2=|\mathcal{D}_{\bal}(i\pi)|^2 \left[\frac{1}{4}\,e^{{\theta}\sinh \theta} \left(2\sinh{\frac{\theta}{2}} \right)^{-2\cosh{\theta}}\right]^{\frac{\alpha}{\pi}}\,,
\eeq 
and for $|\theta|$ large we have
\beq 
|\mathcal{D}_{\bal}(\theta)|^2\sim|\mathcal{D}_{\bal}(i\pi)|^2 \left[\frac{1}{4}\,e^{\frac{\theta}{2}e^\theta} e^{-\frac{\theta}{2}e^\theta} \right]^{\frac{\alpha}{\pi}}= |\mathcal{D}_{\bal}(i\pi)|^2 {4^{-\frac{\alpha}{\pi}}}\,,
\eeq 
which is a constant. This is a radically different asymptotics compared to what we described above. This means that the leading asymptotics of the MFF is dictated by the asymptotics of the MFF of the unperturbed theory, and little can be said about the short-distance scaling of correlators if we do not know the full two-particle form factor. However, from the asymptotics we no longer expect to see a marked difference between the $\alpha$ positive and negative cases. 

There is, however, an important difference in the behaviour of the minimal form factor depending on whether $\alpha>\pi$ or $\alpha<\pi$. In particular, if we consider a $\TTb$ perturbation of the Ising field theory so that
\beqa 
|F_{\rm{min}}^{\rm Ising}(\theta;\alpha)|^2 &=& |F_{\rm{min}}^{\rm Ising}(i\pi;\alpha)|^2 \sinh^2 \frac{\theta}{2} \left[\frac{1}{4}\,e^{{\theta}\sinh \theta} \left|2\sinh{\frac{\theta}{2}} \right|^{-2\cosh{\theta}}\right]^{\frac{\alpha}{\pi}}\nonumber\\
&=& |F_{\rm{min}}(i\pi;\alpha)|^2 4^{-\frac{\alpha}{\pi}}\,e^{\frac{\alpha\theta}{\pi}\sinh \theta} \left|2\sinh{\frac{\theta}{2}} \right|^{2(1- \frac{\alpha}{\pi}\cosh{\theta})}\,, \label{f2}
\eeqa 
we see that for $\alpha<\pi$ the function has a zero at $\theta=0$ while for $\alpha>\pi$ it has an (unphysical) pole at $\theta=0$. We see these properties numerically in Fig.~\ref{fig:mff_theta_various_alpha}. 
The presence of a zero at $\theta=0$ is common with many other IQFTs such as the Ising and sinh-Gordon models. However, the presence of a pole is unphysical and is an indication that the critical value $\alpha_c=\pi$ represents a transition point between two quite different theories.  Interestingly, this kind of transition also occurs for UV-complete models and can be seen for instance when comparing the MFF of the sinh-Gordon model \cite{Fring_1993} with the MFF of the Lee-Yang theory \cite{Z}. Indeed, the $S$-matrices of these two theories are very similar, yet the MFF of the Lee-Yang model with the standard construction would also have a zero at $\theta=0$. A resolution of this problem was put forward in \cite{Z} which consists of multiplying this ``natural" MFF by a function that cancels the pole, while still producing a MFF that is consistent with all axioms and asymptotic requirements. This example provides a pathway for how to deal with the critical values found above. Further discussion will be presented elsewhere \cite{ProcPrag}.

%However, the presence of a pole is unphysical and is an indication that the critical value $\alpha_c=\pi$ represents a transition point between two quite different theories. As noted before, we expect a critical value to arise in interacting models as well. Since in general the MFF is also the full two-particle form factor of the trace of the stress-energy tensor, this means that for $\alpha>\alpha_c$ this form factor will have a pole in the physical strip for which we have no physical interpretation. Furthermore, since the MFF is the building block for the form factors of all fields, this pole may be present for other fields as well, although this will depend on the details of the field and theory.

\section{Conclusion and Outlook}
\label{conclu}
In this paper we revisited the problem of computing the two-particle minimal form factor of an integrable quantum field theory perturbed by $\TTb$. Although the basic structure of this MFF has been identified in previous work \cite{longpaper,Castro-Alvaredo2024compl,ourTTb3,ourboundary} for various types of fields and theories, uniqueness of the solution had not been established. The main result of the present work is to propose two distinct procedures by which the MFF may be fixed unambiguously. We find that the MFF of an IQFT perturbed by $\TTb$ can be uniquely determined by requiring a convergent integral representation, which may be expressed as a regularized version of the usual integral representation. The expression that emerges from this construction has a smooth behaviour in the limit when the perturbation parameter(s) tend to zero and reproduces to the MFF of standard IQFTs (such as the sinh-Gordon model) when the number of irrelevant perturbations is infinite and the couplings are suitably chosen (see Appendix \ref{AppendixB} where the sinh-Gordon example is considered). The formula is
\beq 
F_{\rm min}(\theta;\bal)=F_{\rm min}(\theta) \mathcal{D}_{\bal}(\theta)\,,
\eeq 
where $F_{\rm min}(\theta) $ is the MFF of the unperturbed theory and $\mathcal{D}_{\bal}(\theta)$ is the function (\ref{original}), which is generalised to (\ref{soltwist}) for branch point twist fields \cite{entropy, ourTTb3,theirTTb} and to $\sqrt{\mathcal{D}_{\bal}(2\theta)}$ for the one-particle MFF of boundary IQFTs \cite{ourboundary}. 

A special feature of our solution is that, in the simple case of a single perturbation, say by $\TTb$, there exists a critical value $\alpha=\alpha_c$ above which the MFF develops a pole of order $\frac{\alpha}{\pi}-\kappa$ at $\theta=0$, where $\kappa$ is a property of the unperturbed MFF. The existence of a ``transition'' in the physical properties of the model seems reminiscent of what is observed in the thermodynamic Bethe ansatz analysis (see e.g. \cite{Cavaglia:2016oda, LeClair_2021}), in particular the emergence of a Hagedorn transition at finite temperature. However, there are notable differences between the two situations and we think that they are  not related. For one, the Hagedorn transition occurs for \emph{any} negative value of $\alpha$ (with our prescriptions), whereas for our MFF the ``transition'' happens at a specific, positive value of $\alpha$. Secondly, it is well known that the Hagedorn transition is particular to the pure $\TTb$ perturbation and absent for higher spins (unless one modifies the driving term of the TBA and considers specific generalized Gibbs ensembles, as discussed in \cite{RMO}). In our case, however, the critical point is present for any finite number of perturbations. 
Perhaps, the most interesting observation is that, as discussed at the end of Section~\ref{correlation} the presence of a pole at $\theta=0$ is a phenomenon that also occurs for some UV-complete theories when their MFFs are constructed via the standard integral representation discussed in this paper. The simplest example is provided by the Lee-Yang theory whose MFF was constructed in \cite{Z}. In this case the naturally occurring singularity at $\theta=0$ is cancelled out by simply modifying the MFF by a multiplicative factor which has a zero at $\theta=0$. We think therefore that it may still be possible to make sense of solutions with $\alpha>\alpha_c$ by a similar modification of the MFF. How this factor should be chosen for IQFTs perturbed by a finite number of irrelevant perturbations is however an issue that needs further discussion. A more detailed discussion can be found in our upcoming work \cite{ProcPrag}.

Although our analysis provides a way to fix the MFF which is reminiscent of the traditional construction of Karowski-Weisz \cite{KarowskiWeisz:1978vz}, it does not make the construction of higher particle solutions any easier. In fact, the correct prescription for computing such solutions is still not fully understood. Solving by plugging our new MFF into the standard ansatz for models with a single particle (like sinh-Gordon \cite{Fring_1993}) and imposing factorization (as we previously did) leads to solutions which are not only plagued by square-root factors (thus, points of nonanalyticity), similar to \cite{longpaper,Castro-Alvaredo2024compl}, but also lead to a partial cancellation of the MFF in the form factor ansatz (this is discussed in some detail in Appendix \ref{AppendixB}). This second issue was not present in our earlier work because we had set the function $C_{\bel}(\theta)=1$ in (\ref{minimal}) and worked with the simplest possible form of the MFF. 

This cancellation suggests that it may not be possible for the kinematic residue equation as it is, the standard form factor ansatz with our complete MFF, and the assumption of factorization into a $\bal$-dependent and an undeformed part, to all hold simultaneously. As we already discussed in the introduction, the locality properties of the observables are crucial, and specifically the kinematic residue equation is underpinned by the notion of locality (or semi-locality). The work \cite{Bostelmann:2014zda} sheds some light on the validity of the standard form factor program. Using formal algebraic techniques, local observables for IQFTs are built by infinite expansions whose coefficients are a sort of generalization of the usual form factors, coinciding with them for certain regions of the arguments. Local observables are characterized by the analyticity properties of their expansion coefficients. These coefficients must satisfy a set of axioms that correspond exactly to the usual form factor equations. Interestingly, in the conclusion of \cite{Bostelmann:2014zda} it is pointed out how this result could be ``employed to prove \textit{non}-existence of local observables in the sense of a \textit{no-go theorem}'' by showing that for some $S$-matrices it is impossible to satisfy all the ``axioms".

The CDD-factor \eqref{sum} is proposed as an example of such an $S$-matrix. Indeed the \textit{non}-existence of local observables in these theories was widely conjectured \cite{GrosseGandalf_2007,Dubovsky:2012wk}, based mainly on the failure of known algebraic and form factor methods, but this has never been rigorously established. It would be interesting to understand if our constructive approach can be used to prove this statement more rigorously. 

As shown in Appendix \ref{AppendixB} a factorised ansatz, where the form factor is a product of the unperturbed solution times a function of the perturbation parameter, seems extremely natural for $\TTb$-perturbed theories but is easily shown not to produce the correct solutions when considering more standard IQFTs, that is, when considering the case of infinitely many irrelevant perturbations. We have checked this explicitly for the sinh-Gordon model, where a factorized ansatz gives solutions that are distinct from those known in the literature \cite{Fring_1993,1993:Koubek,freefield1} and contain square roots, as already discussed  (see Appendix \ref{AppendixB}). 

Regarding future work, there are at least two avenues opened to us: it may be that there are non-factorized solutions that we have just not been able to construct by the usual methods, would probably be highly non-trivial, and are still to be found, or, the standard form factor program and in particular the kinematic residue equation, is not valid in its usual form for $\TTb$-perturbed theories. The latter is a very plausible suggestion since giving up the notion of UV completion and locality does away with two of the basic underlying principles of the form factor program as originally formulated.  Our current work, as well as \cite{longpaper,Castro-Alvaredo2024compl}, provides a very strong indication that the standard FF program may not admit analytic solutions for theories perturbed by $\TTb$. Even if this is the case, it would be interesting to understand how the breakdown of the program takes place (by using perturbative methods, for instance). 

\medskip
\noindent {\bf Acknowledgments:} Fabio Sailis is grateful to the School of Science and Technology of City St George's, University of London for a PhD studentship. Olalla A. Castro-Alvaredo and Fabio Sailis are grateful for the opportunity to discuss parts of this work at various recent conferences and seminars, including at SISSA (Italy), May (2025), the Student Workshop on Integrability, Budapest (Hungary), June (2025), the XVI International Workshop Lie Theory and Its Applications in Physics, Varna (Bulgaria), June (2025), and the XXIXth International Conference on Integrable Systems and Quantum Symmetries, in Prague (Czech Republic), July (2025). 
\appendix

\section{Revisiting the Sinh-Gordon Model}
\label{AppendixB}

\subsection{Factorizing the Kinematical Pole}

The sinh-Gordon model is the simplest IQFT which is also interacting, hence has a non-trivial scattering matrix \cite{SSG,SSG2,SSG3}. The two-body scattering matrix takes the simple form
\beq 
S_{\bal}(\theta)=\frac{\tanh\frac{1}{2}\left(\theta-\frac{i\pi B}{2}\right)}{\tanh\frac{1}{2}\left(\theta+\frac{i\pi B}{2}\right)}\,,
\label{SmatrixSinhGordon}
\eeq 
where $B$ is a coupling constant that takes values between 0 and 2. The value $B=1$ is known as the self-dual point. Following the notation in our introduction we have called this $S$-matrix $S_{\bal}(\theta)$. The reason for this is that there is an alternative way of thinking about this function, namely as an $S$-matrix of the type (\ref{Smatrix})  where $S(\theta)=-1$, corresponding to the Ising field theory (Majorana fermion) and 
\beqa
\Phi_{\bal}^{\rm shG}(\theta)&=&-\frac{\tanh\frac{1}{2}\left(\theta-\frac{i\pi B}{2}\right)}{\tanh\frac{1}{2}\left(\theta+\frac{i\pi B}{2}\right)}= \frac{\sinh\theta-i \cos \frac{b\pi}{2}}{\sinh\theta+i \cos \frac{b\pi}{2}}\nonumber\\
&=& \exp\left[-4 i \sum_{s \, \in\, 2\mathbb{Z}^{+}-1} i^{s+1}\frac{\cos\frac{s\pi b}{2}}{s} \sinh(s\theta) \right]\,,
    \label{funnyS}
\eeqa
where $b:=B-1$. This means in particular that
\beq 
\alpha_{2n-1}= \frac{4 (-1)^{n}}{2n-1}\,{\cos\frac{(2n-1)\pi b}{2}}\,,
\eeq 
and plugging these values into (\ref{original}) provides yet another way to represent the MFF of the sinh-Gordon theory:
\beqa 
&&\frac{F^{\rm shG}_{\rm min}(\theta;\bal)}{F^{\rm shG}_{\rm min}(i\pi;\bal)}=-i \sinh\frac{\theta}{2} \\
&& \!\!\times \prod_{n\in \mathbb{Z}^+}\left[-\frac{1}{2}\,e^{\frac{\theta-i\pi}{2}\sinh((2n-1)\theta)-\sum\limits_{m=1}^{2n-2}\frac{\cosh(m\theta)}{2n-1-m}+\ c_{2n-1}} \left[2i \sinh{\frac{\theta}{2}} \right]^{-\cosh{((2n-1)\theta)}}\right]^{{\frac{4 (-1)^{n} \cos\frac{(2n-1)\pi b}{2}}{\pi(2n-1)}}}\,.\nonumber
\eeqa 
Employing this representation of the sinh-Gordon scattering matrix, it is possible to show that its MFF \cite{MMF} admits a representation of the type (\ref{minimal}) where all values of $\bal$ and $\bel$ are fixed by the requirements of analyticity and asymptotics. The exact formula for the function $C_{\bel}^{\rm shG}(\theta)$ was obtained in \cite{MMF}. The explicit form of this function is not needed for our current purposes. An important property of the MFF of sinh-Gordon is that 
\beq
F_{\rm min}^{\rm shG}(\theta; \bal)F_{\rm min}^{\rm shG}(\theta + i \pi; \bal)= \frac{\sinh\theta}{\sinh
\theta+ i \cos \frac{\pi b}{2}}\,.
\label{normalkinpol}
\eeq
Our aim is to find higher-particle form factors of the sinh-Gordon model by employing a factorised ansatz. We will start by separating the Ising, $\bal$ and $\bel(\bal)$ parts, using
$F_{\rm min}^{\rm Ising}(\theta)=- i \sinh\frac{\theta}{2}$ and\footnote{From \cite{MMF} it is interesting to note that only the terms involving $\beta_{s}$ with $s$ even contribute to obtaining the property \eqref{normalkinpol}. This excludes all terms involving dilogarithm functions. Note also that compared to \cite{MMF} we included an extra multiplicative factor $\sqrt{2}$ into the definition of $C_{\bel}^{\rm shG}(\theta)$.} 
\beq 
C_{\bel}^{\rm shG}(\theta) C_{\bel}^{\rm shG}(\theta+i\pi)= \frac{2}{\sqrt{\sinh^2{\theta} + \cos^2{\frac{\pi b}{2}}}}=-\frac{2i}{\sqrt{\Psi^{\rm shG}(\theta)\Psi^{\rm shG}(-\theta)}}\,, 
\label{CC}
\eeq 
\beq 
\varphi_{\bal}^{\rm shG}(\theta)\varphi_{\bal}^{\rm shG}(\theta+i\pi)=\sqrt{\Phi_{\bal}^{\rm shG}(\theta)}= \sqrt{-\frac{\sinh\theta-i \cos \frac{\pi b}{2}}{\sinh
\theta+ i \cos \frac{\pi b}{2}}}\,,
\eeq 
where
\beq 
\Psi^{\rm shG}(\theta)=\sinh\theta+i \cos\frac{\pi b}{2}\,,
\label{psiShG}
\eeq 
with
\beq 
\Psi^{\rm shG}(\theta) \Phi_{\bal}^{\rm shG}(\theta)=\Psi^{\rm shG}(-\theta)\,.
\label{psiexchange2}
\eeq 
It is well known that the formula \eqref{normalkinpol} plays a critical role in determining the higher-particle form factors. Writing it in the factorised form above  allows us to separate the contributions related to the $\bal$ parameters, $\bel=\bel(\bal)$ parameters  and the unperturbed theory, parameters which are inherited from the representation (\ref{betas}). We will now ``track" these separate contributions into the structure of higher-particle form factors and investigate what they reveal about the form factors of $\TTb$-perturbed theories.

\subsection{$\TTb$-Like Solution of the sinh-Gordon Model}
Typically, the form factors of a field $\mathcal{O}$, defined in (\ref{introFF}) have the structure
\beq
F^{\mathcal{O}}_n(\theta_1,\ldots,\theta_n; \bal)=H^{\mathcal{O}}_n(\bal) Q^{\mathcal{O}}_n(\theta_1,\ldots,\theta_n; {\bal}) \prod_{i<j} \frac{F_{\rm min}(\theta_{ij};{\bal})}{e^{\theta_i}+e^{\theta_j}}\,,\label{ansatz}
\eeq
where 
\beqa
Q^{\mathcal{O}}_{n+2}(\theta+i\pi, \theta,\theta_1,\ldots,\theta_{n};\boldsymbol{\alpha})=G_n(\theta, \theta_1,\ldots, \theta_n;\bal,\bel)  Q^{\mathcal{O}}_n (\theta_1,\ldots, \theta_n;\boldsymbol{\alpha})\,.
\eeqa
In the sinh-Gordon case we have the structure (normalisation constants can be absorved into the equation for the constants $H^{\mathcal{O}}_n(\bal)$)
\beqa 
G_n(\theta, \theta_1,\ldots, \theta_n;\bal) &:=& i^n e^{(n+1)\theta} \prod_{j=1}^n e^{\theta_j} \left( \prod_{j=1}^n  \sqrt{\Psi^{\rm shG}(\theta-\theta_j)\Psi^{\rm shG}(\theta_j-\theta)}\right)
\nonumber\\
&& \times \left[\prod_{j=1}^n \Phi^{\rm shG}_{\boldsymbol{\alpha}}(\theta-\theta_j)^{-\frac{1}{2}} -\gamma_{\mathcal{O}} (-1)^n \prod_{j=1}^n \Phi^{\rm shG}_{\boldsymbol{\alpha}}(\theta-\theta_j)^{\frac{1}{2}}\right]\,.
\label{Gn1}
\eeqa 
Employing the property \eqref{psiexchange2} and the fact that $\Phi_{\bal}(\theta)=(\Phi_{\bal}(-\theta))^{-1}$ (by unitarity) we can rewrite
\beqa 
G_n(\theta, \theta_1,\ldots, \theta_n;\bal,\bel) = i^n e^{(n+1)\theta} \prod_{j=1}^n e^{\theta_j} \left[\prod_{j=1}^n \Psi^{\rm shG}(\theta-\theta_j) -\gamma_{\mathcal{O}} (-1)^n \prod_{j=1}^n \Psi^{\rm shG}(\theta_j-\theta)\right]\,.
\label{Gn2}
\eeqa 
These equations are equivalent to those presented in \cite{Fring_1993} and \cite{1993:Koubek}. Proceeding in this way allows us also to see how the square roots present in the form \eqref{Gn1} are absent when combining the $\bal$ and $\bel$ factors together in the sinh-Gordon model case. If we stick with the representation (\ref{Gn1}) though we have a neat separation into three different contributions: the factor $ e^{(n+1)\theta} \prod_{j=1}^n e^{\theta_j}$ is the contribution from the Ising field theory, the next factor incorporates the $\bel$ dependence and the factor in the second line incorporates the $\bal$ dependence and, as expected, is of exactly the type found for generalised $\TTb$-perturbed theories \cite{longpaper,ourTTb3}. Mixing the $\bal$ and $\bel$ factors together with the Ising factors gives the standard solutions for the sinh-Gordon operator content. We will see in the following that instead, imposing factorization we end up with different solutions even though the S-matrix is still the sinh-Gordon one. 

\subsection{A Factorized Solution}

Therefore, it is natural to try to find solutions to these equations using the same methodology as in \cite{longpaper,ourTTb3}. We make a solution ansatz stating that the sinh-Gordon form factors of some "local" field factorize as the form factors of an Ising field theory times a $\bal$-dependent part. Furthermore, contrary to our treatment in \cite{longpaper,ourTTb3} where we considered the case with all $\beta_i=0$, in the sinh-Gordon model we have parameters $\beta_i$ entering the MFF, so we need a more general ansatz. Again, we assume further factorisation into an $\bal$-dependent and $\bel$-dependent part. We write
\beq
Q_n^{\mathcal{O}}(\theta_1,\ldots,\theta_n;\bal) = Q_n^{\mathcal{O}}(\theta_1,\ldots,\theta_n) \Theta_n^{\mathcal{O}}(\theta_1,\ldots,\theta_n;{\bal})\Xi_n^{\mathcal{O}}(\theta_1,\ldots,\theta_n;\boldsymbol{\beta}(\bal))\;,
\label{Qfactorization}
\eeq
Where 
\beq 
Q_{n+2}^{\mathcal{O}}(\theta+i\pi,\theta,\theta_1,\ldots,\theta_n)=e^{(n+1)\theta} \left[\prod_{j=1}^n e^{\theta_j} \right] Q_n(\theta_1,\ldots,\theta_n)\,,
\label{QIsing}
\eeq 
which (up to normalisation) is the equation for the Ising form factors of the order field $\mu$ (for $n$ even) and for the disorder field $\sigma$ (for $n$ odd) \cite{Yurov:1990}. The solution to this equation, or rather, the full form factors are very simple and proportional to he product $\prod_{i<j} \tanh\frac{\theta_{ij}}{2}$ which automatically incorporates the denominator $e^{\theta_i}+e^{\theta_j}$ in  \eqref{ansatz}. The next equation of interest is
\beq 
\Theta_{n+2}^{\mathcal{O}}(\theta+i\pi,\theta,\theta_1,\ldots,\theta_n;\bal)=\left[\prod_{j=1}^n \Phi_{{\alpha}}(\theta-\theta_j)^{-\frac{1}{2}} -\gamma_{\mathcal{O}} (-1)^n \prod_{j=1}^n \Phi_{{\alpha}}(\theta-\theta_j)^{\frac{1}{2}}\right]\Theta_{n}^{\mathcal{O}}(\theta_1,\ldots,\theta_n; \bal)\,,
\label{Tsinh}
\eeq 
which satisfies the generalised $\TTb$ form factor equation discussed and solved in \cite{longpaper,ourTTb3}. The general solution reads 
\beqa 
    \Theta_n^{\mathcal{O}}(\theta_1,\ldots,\theta_n;{\bal}) = \prod_{i=1}^n \sqrt{\frac{\prod\limits_{j=1}^n S_{\boldsymbol{\alpha}}(\theta_{ij})^{1/2} - \gamma^{\mathcal{O}} \prod\limits_{j=1}^n S_{\boldsymbol{\alpha}}(\theta_{ij})^{-1/2}}{\prod\limits_{j=1}^n S(\theta_{ij})^{1/2} - \gamma^{\mathcal{O}} \prod\limits_{j=1}^n S(\theta_{ij} )^{-1/2}}}\,,
    \label{solutionTheta}
\eeqa 
as shown in \cite{Castro-Alvaredo2024compl}. The $\bel$-dependent recursive equation instead reads
\beq
\Xi^{\mathcal{O}}_{n+2}(\theta+i\pi,\theta,\theta_1,\ldots,\theta_{n};\boldsymbol{\beta}(\bal))= i^n \left[\prod_{j=1}^n \sqrt{\Psi^{\rm shG}{(\theta-\theta_j)}\Psi^{\rm shG}{(\theta_j-\theta)}}\right]
\Xi_n^{\mathcal{O}}(\theta_1,\ldots,\theta_n;\boldsymbol{\beta}(\bal))\,.
\label{XSinh}
\eeq
Like in the Ising model and in its deformed version,  we expect to have different sectors dictated by the $\mathbb{Z}_2$ symmetry: either ``even" fields such as the order field $\mu$ and the energy field $\varepsilon$ and ``odd" fields, such as $\sigma$. Their counterparts in the sinh-Gordon model, which also possesses this symmetry, should at least be in the same symmetry sector, with the fundamental sinh-Gordon field $\phi$ in the ``odd" sector and the trace of the stress-energy tensor in the ``even" sector. Anyway, we notice that the factor of local commutativity do not enters equation \eqref{XSinh}. A solution to this equation is relatively easy to find, namely
\beq
\Xi_n^{\mathcal{O}}(\theta_1,\ldots,\theta_n;\boldsymbol{\beta}(\bal)) = {v}_n \prod_{i=1}^n \prod_{j=1}^n \sqrt[4]{\Psi^{\rm shG}(\theta_{ij})}\,,
\label{psi}
\eeq
where 
$v_n$ is a normalisation constant which satisfies
\beq 
v_{n+2}=\cos\frac{\pi b}{2} \, v_n\,,
\eeq 
which, depending on whether $n$ is even or odd gives solutions
\beq 
v_{2k}= \left(\cos\frac{\pi b}{2}\right)^{k} v_0 \quad  \mathrm{and}\quad
v_{2k+1}= \left(\cos\frac{\pi b}{2}\right)^{k-1} v_1\,. 
\eeq 
In addition to this, if we specialise to $n=2k+1$ odd and $\gamma_{\mathcal{O}}=1$, we get for the $\bal$-dependent part 
\beq
\Theta_{2k+1}^{\mathcal{O}}(\theta_1,\ldots,\theta_{2k+1};\boldsymbol{\alpha}) = 
\frac{1}{2^k \sqrt{2}} \prod_{i=1}^{2k+1}  \sqrt{\prod\limits_{j=1}^{2k+1} \Phi^{\rm shG}_{\boldsymbol{\alpha}}(\theta_{ij})^{1/2} + \prod\limits_{j=1}^{2k+1} \Phi^{\rm shG}_{\boldsymbol{\alpha}}(\theta_{ij})^{-1/2}}\,,
\eeq
and similarly for $n$ even and $\gamma_{\mathcal{O}}=-1$. This expression is rather complicated and there is no immediately obvious way to simplify it. In our example, we have in addition that $\Phi^{\rm shG}_{\bal}(\theta)\Phi^{\rm shG}_{\bal}(-\theta)=1$ so we can rewrite the formula in several equivalent ways. For example,
\beqa
\prod_{i=1}^{n}  \sqrt{\prod\limits_{j=1}^{n} \Phi^{\rm shG}_{\boldsymbol{\alpha}}(\theta_{ij})^{1/2} + \prod\limits_{j=1}^{n} \Phi^{\rm shG}_{\boldsymbol{\alpha}}(\theta_{ji})^{1/2}}&=&\prod_{i=1}^{n}\sqrt{1 + \prod\limits_{j=1}^{n} \Phi^{\rm shG}_{\boldsymbol{\alpha}}(\theta_{ji})}\,.
\label{reps}
\eeqa
It seems therefore that it is possible to consistently solve the form factor equations by assuming factorisation, but the resulting solutions are distinct from those previously found and contain square roots. 
\subsection{Another Solution}

We could instead assume factorisation but separate only the Ising part from the rest. That is, instead of (\ref{Qfactorization}) we now have 
\beq
Q_n^{\mathcal{O}}(\theta_1,\ldots,\theta_n;\bal) = Q_n^{\mathcal{O}}(\theta_1,\ldots,\theta_n) \Gamma_n^{\mathcal{O}}(\theta_1,\ldots,\theta_n;{\bal})\;,
\label{Qfactorization3}
\eeq
If we specialise to the sector $\gamma_{\mu}=-1$ and with a sum over even particle numbers, we get
 \beq
 \Gamma^{\mathcal{\mu}}_{2n+2}(\theta+i\pi,\theta,\theta_1,\ldots,\theta_{2n};\boldsymbol{\alpha})=  \left[\prod_{j=1}^{2n} \Psi^{\rm shG}(\theta-\theta_j) +  \prod_{j=1}^{2n} \Psi^{\rm shG}(\theta_j-\theta)\right] \Gamma_{2n}^{\mathcal{\mu}}(\theta_1,\ldots,\theta_{2n};\boldsymbol{\alpha})\,.
 \label{GammaKineq}
\eeq
To solve this equation, we just need the property $\Psi^{\rm shG}(\theta_{ij} + i \pi)=\Psi^{\rm shG}(\theta_{ji})$ and we find the solution
\beq
\Gamma_{2n}^{\mathcal{\mu}}(\theta_1,\ldots,\theta_{2n};\boldsymbol{\alpha}) = C_{2n} \frac{ \prod_{i=1}^{2n} \sqrt{\prod_{j=1}^{2n} \Psi^{\rm shG}(\theta_{ij}) +  \prod_{j=1}^{2n} \Psi^{\rm shG}(\theta_{ji})}}{\prod_{i<j} \sqrt[4]{ \Psi^{\rm shG}(\theta_{ij}) \Psi^{\rm shG}(\theta_{ji})}}\,.
\label{solutionGamma2}
\eeq
with $C_{2n}$ a constant. Once more the solution involves roots which do not cancel out and therefore it is distinct from the usual solutions for the sinh-Gordon model. In addition, the factor in the denominator can be shown to cancel part of the contribution coming from the product of MFFs. It seems clear that the Ising part of the MFF and the rest must be considered together in order to solve the kinematic residue equation. 

\section{Generalisation to Branch Point Twist Fields}
\label{BPTFs}
In this Appendix we present a generalisation of the results of Subsection \ref{solution} to the case of branch point twist fields \cite{entropy,ourTTb3}. For our purposes, the starting point is the integral representation
\beq
\log{F_{\rm min}(\theta;n)} =    \frac{1}{i \pi n} \cosh^2{\frac{\theta}{2n}} \int_0^{\infty} dt \frac{ \tanh \frac{t}{2n} \ln{S(t)}}{ \cosh{\frac{t}{n}} - \cosh{\frac{\theta}{n}} }  \,, \qquad \textrm{for}\; 0 \leq {\rm Im}(\theta) \leq 2 \pi n \,,
\eeq
which reduces to (\ref{NiedermeierMFF}) for $n=1$.  Again, we can obtain this representation by inverting the Fourier transform of the $S$-matrix, that is, by expressing the kernel $g(t)$ in terms of the Fourier transform of $\ln S(t)$ in the usual representation (see \cite{entropy}). As we did earlier, let us consider the deformation of the MFF resulting from a generalised $\TTb$ perturbation in its regularised form
\beq
\log{\mathcal{D}_{\bal}(\theta;n)}= \frac{1}{i \pi n} \cosh^2{\frac{\theta}{2n}} \int_0^{\infty} dt \frac{ \tanh\frac{t}{2n} }{\cosh{\frac{t}{n}} - \cosh{\frac{\theta}{n}} } \left(\ln{\Phi_{\bal}(t)}-\partial_t \ln \hat{\Phi}_{\bal}(\theta)\right)\,,
\label{Dreg}
\eeq
where $\hat{\Phi}_{\bal}(\theta)$ is the function defined in (\ref{phihat}). Expanding for small $|\theta|$ we can write
\beq
\frac{\cosh^2{\frac{\theta}{2n}}\tanh{\frac{t}{2n}}}{\cosh{\frac{t}{n}}-\cosh{\frac{\theta}{n}}}= \sum_{m=  1}^{\infty} \left[ \cosh{\frac{k \theta}{n}}+(-1)^{m+1} \right] e^{-m \frac{t}{n}}\,,
\eeq
which when substituting in \eqref{Dreg} gives the simple integral 
\beq
 \int_0^{\infty} dt  e^{-t (\frac{m}{n} +s)} = -\frac{1}{\frac{m}{n}+s}\,.
\eeq
The resulting contribution for a given spin $s$ is
\beq 
\log{\mathcal{D}_{\alpha_s}(\theta;n)}=\frac{\alpha_s}{\pi} \sum_{m=1}^{\infty} \frac{\cosh{\frac{m \theta}{n}}+(-1)^{m+1} }{m+s \,n}\,.
\eeq 
The sum can be performed as before and it gives a solution, with the structure found in \cite{ourTTb3,theirTTb}, namely
\beqa 
\sum_{m = 1}^{\infty} \frac{\cosh{\frac{m \theta}{n}}}{m+s n} =\frac{\theta-i\pi n}{2n} \sinh(s \theta) -\frac{1}{2}\log\left(-4\sinh^2\frac{\theta}{2n}\right) \cosh(s\theta) -\sum_{m=0}^{n s-1} \frac{1}{n s-m}\cosh\frac{m\theta}{n}\,,
\eeqa
and the constant is
\beqa 
\sum_{m=1}^\infty \frac{(-1)^{m+1}}{m + s n}&=& \sum_{m=sn+1}^\infty \frac{(-1)^{m+ s n + 1}}{m}= (-1)^{s n}\sum_{m= 1}^\infty \frac{(-1)^{m+1}}{m} + \frac{1}{sn} -(-1)^{sn} c_{sn}\nonumber\\
&=& (-s)^{sn} \log 2 + \frac{1}{sn} -(-1)^{sn} c_{sn}\,.
\eeqa 
where $c_{sn}$ is the constant defined in (\ref{cn}). In summary, normalising $\mathcal{D}_{\bal}(\theta;n)$ by its value at $\theta=i\pi n$ we can write
\beq 
\frac{\mathcal{D}_{\bal}(\theta;n)}{\mathcal{D}_{\bal}(i\pi n; n)}= \prod_{s \in 2\mathbb{Z}^{+}-1}\left[-\frac{1}{2^{(-1)^{s n}}}\,e^{\frac{\theta-i\pi n}{2 n}\sinh(s\theta)-\sum\limits_{m=1}^{n s-1}\frac{\cosh\frac{m\theta}{n}}{n s-m} + (-1)^{n s}{c}_{s n}} \left(2i \sinh{\frac{\theta}{2 n}} \right)^{-\cosh{(s\theta)}}\right]^{\frac{\alpha_s}{\pi}}\,.
\label{soltwist}
\eeq 
Setting $n=1$ we recover the solution (\ref{original}). The formulae above can be easily generalized to real $n$ (as opposed to $n$ integer) by introducing integer part symbols in some of the summation limits. 
\bibliography{bibliography}
\end{document}